\documentclass[10pt]{article}%

\usepackage[a4paper,top=2.5cm,left=2cm,bottom=2.5cm,right=2cm]{geometry}
\usepackage{amsmath}
\usepackage{amssymb}
\usepackage{amsthm} 
\usepackage{mathabx} 
\usepackage{graphicx}
\usepackage{hyperref} 
\usepackage[authoryear,sort&compress]{natbib} 
\usepackage{subfigure}  
\usepackage{color}
\usepackage{ulem} 

\newcommand{\apj}{The Astrophysical Journal}

\newcommand{\supno}{Note}

\newtheorem{proposition}{Proposition} 
\newtheorem{lemma}{Lemma} 
\newtheorem{corollary}{Corollary} 

\theoremstyle{remark}
\newtheorem{note}{Note}

\usepackage{lineno}

\begin{document}
\title{The power-law distribution in the geometrically growing system: Statistic of the COVID-19 pandemic}

\author{Kim Chol-jun  \\%
\small \textit{Department of Astronomy, Faculty of Physics, \textbf{Kim Il Sung} University, DPR Korea }
\\
\small postal code: +850      
\\
\small email address: cj.kim@ryongnamsan.edu.kp
}
\date{\small December 19, 2020, Rev. Jan 15, May 9, Sep 23, 2021}
\maketitle
\begin{abstract}
The power-law distribution is ubiquitous and its mechanism seems to be various. We find a general mechanism for the distribution. The distribution of a geometrically growing system can be approximated by a log - completely squared chi distribution with 1 degree of freedom (log-CS$\chi_1$), which reaches asymptotically a power-law distribution, or by a log-normal distribution, which has an infinite asymptotic slope, at the upper limit. For the log-CS$\chi_1$, the asymptotic exponent of the power-law or the slope in a log-log diagram seems to be related only to the variances of the system parameters and their mutual correlation but independent of an initial distribution of the system or any mean value of parameters. We can take the log-CS$\chi_1$ as a unique approximation when the system should have a singular initial distribution. The mechanism shows a comprehensiveness to be applicable to wide practice. We derive a simple formula for the Zipf's exponent, which will probably demand that the exponent should be near -1 rather than exactly -1. We show that this approach can explain statistics of the COVID-19 pandemic.
\end{abstract}


\textbf{While we analyzed the power-law distribution of the cosmic ray spectrum, we referred to even other sciences such as economics to search for a universal statistical mechanism for the power-law spectrum. However, we found that every existing mechanism accompanies so complex assumptions of local circumstance that they are less applicable to other subjects. On the basis of the analysis, we built a comprehensive and comprehensible model of power-law distribution which is able to be shared in the astrophysics, economics and other sciences. We can claim that the variances in the age of growth and growth rate and their correlation can lead to a skewed distribution of power law.}

\section{Introduction}\label{sec:intro}

The power-law distribution is ubiquitous. We can find examples related to the distribution in the universe such as the cosmic ray (CR) spectrum and Salpeter's initial mass function (IMF) of stars \citep{Salpeter1955}. Kolmogorov spectrum in turbulence \citep{Kolmogorov1941, Kolmogorov1963} is another example in physics. However, the dynamic interpretation of them is not easy. For example, the analytic interpretation of the Kolmogorov spectrum with Navier-Stokes equation seems almost impossible and a dimensional analysis is the most we can do now. 

The power-law spectrum appears in the world around us. Pareto and Zipf's leading works \citep{Pareto1896, Zipf1932, Zipf1949} have shown several phenomena having the power-law distribution such as the income, word frequency in a novel and the population in city but they also have no perfect solutions yet. To interpret the power-law distribution for the problems, several mechanisms have been proposed (refer to \citet{Newman2005}, \citet{Sornette2004} or \citet{Mitzenmacher2004} for detailed introduction of mechanisms). 

Those mechanisms have shown many possibilities to reach the power-law distribution. However, they have drawbacks as well, some of them we describe here. First, they are less transplantable. Most mechanisms supposed so specific processes and conditions that they could be hardly adopted in other phenomena or even in the same phenomena with another condition. The preferential attachment or Yule-Simon process \citep{Yule1924, Simon1955} to explain the distribution of species and genera in nature or the population in city can not or have never used to explain, for instance, the CR acceleration, in spite of that those phenomena have a similarity in essence as we will mention below. Even as regards the two modes of Fermi acceleration of CR, we can not find formula for asymptotic slope of CR spectrum in the case of second-order Fermi mode while that for the first-order mode is given in references \citep[e.g., see][]{Gaisser1990}. \cite{Gabaix1999a} showed that the Zipf's law for the population in cities may result on stochastic way. His theory depends on a supposition for the steadiness of normalized distribution, which however does not hold in case of coronavirus as we will see in Sec.~\ref{sec:covid}. \citet{Reed2004} supposed an exponential distribution of the observation time, which is a novel supposition on the time variance, and obtained a ``double - Pareto distribution,'' but his supposition seems less practical at least in the case of CR.

Secondly, most of previous works tried to generate the power-law distribution itself. Many such practical phenomena, however, accompany a roll-over, i.e. a convex lower part of distribution implying a decreasing frequency in the lower sizes. This fact seems to have another mechanism apart from the power law. Previous workers introduced some very specific and complex astrophysical or economical reasoning to explain the rollover and to join it with the power law smoothly, which made the theories bulky and illegible. For example, in order to explain a roll-over in IMF they had to devise a lower-mass limit of star formation \citep[e.g., see][]{Chabrier2003}. For the population in cities \cite{Gabaix1999a, Gabaix1999b} tried to demonstrate ``too few'' small cities via a economic and psychological model of migration called ``reflected geometric Brownian motion.''  

In this paper we try to find another mechanism for the power-law distribution. We can find that aforementioned some phenomena are similarly related to the growth, especially a geometric growth, i.e. proportional to the current size. First, we consider an example of the distribution for the CR acceleration in Sec.~\ref{sec:numsim}. Then we establish a general formalism for distribution in a geometrically growing system in Sec.~\ref{sec:formal}. And we consider statistics of the pandemic as an application in Sec.~\ref{sec:covid}.

\section{Numerical simulation}\label{sec:numsim}
We start with a simulation of the particle acceleration at the shock of a fixed radius around a celestial object. In fact, the particles can be accelerated via any shock it encounters. The energy of a particle grows geometrically at each encounter (\supno~\ref{note:CRaccelintro}). Here we do not try any physical reasoning or verification for a specific acceleration mode but pursue only a possibility of acceleration to reach the power-law distribution. In the simulation we adopt the first-order Fermi acceleration, which causes the energy increment proportional to the shock velocity. Additionally, we suppose that the incident angle of a particle to the shock front is equal to the reflection angle, i.e. the mirror reflection, which differs from the diffusive scattering as in the Fermi accelerations. To enhance the encountering numbers of particles to the shock front, we apply a periodical boundary condition to the system where a particle escaped from one side of the rectangular boundary of system is entering the opposite side. 

\begin{figure*}
\centering
\begin{tabular}[t]{cc}
\raisebox{-17ex}[0ex][0ex]{\subfigure[]{\includegraphics[width=0.25\textwidth]{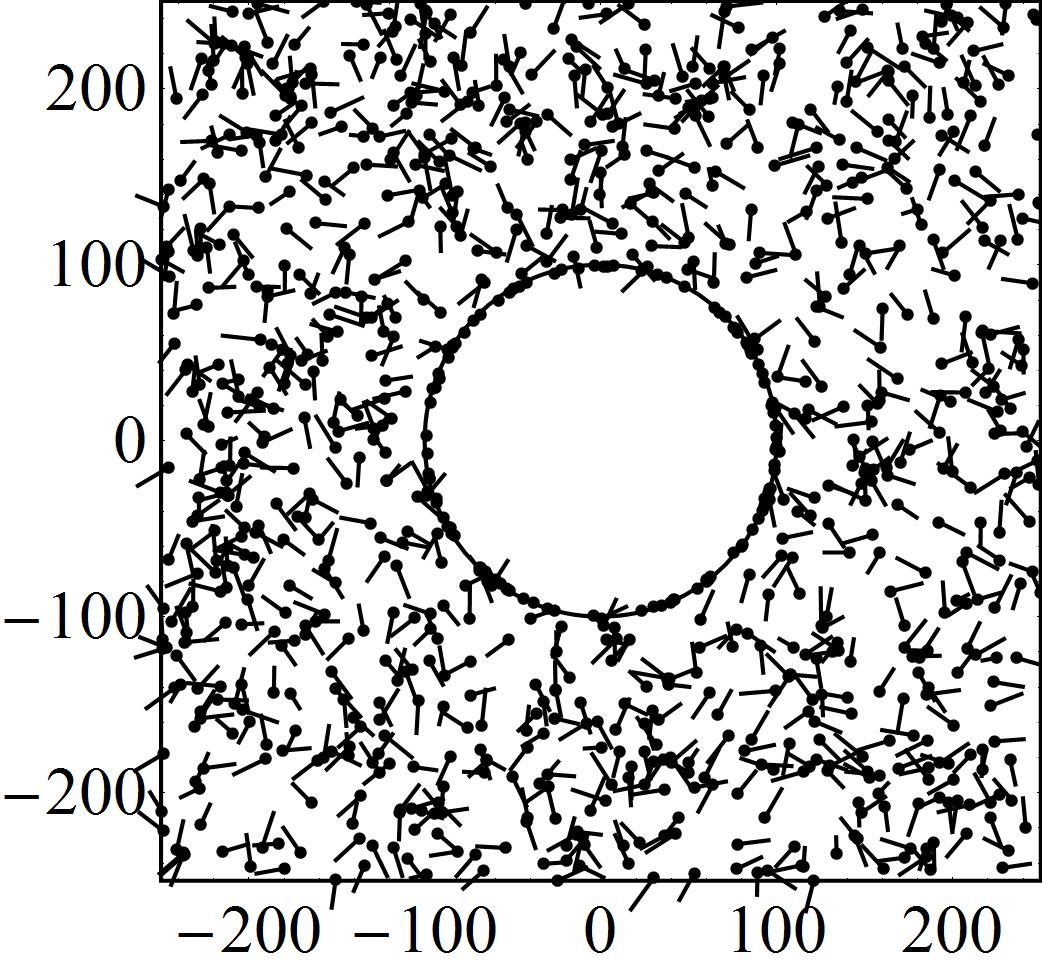}\label{fig:Configa}}}&
\subfigure[]{\includegraphics[width=0.3\textwidth]{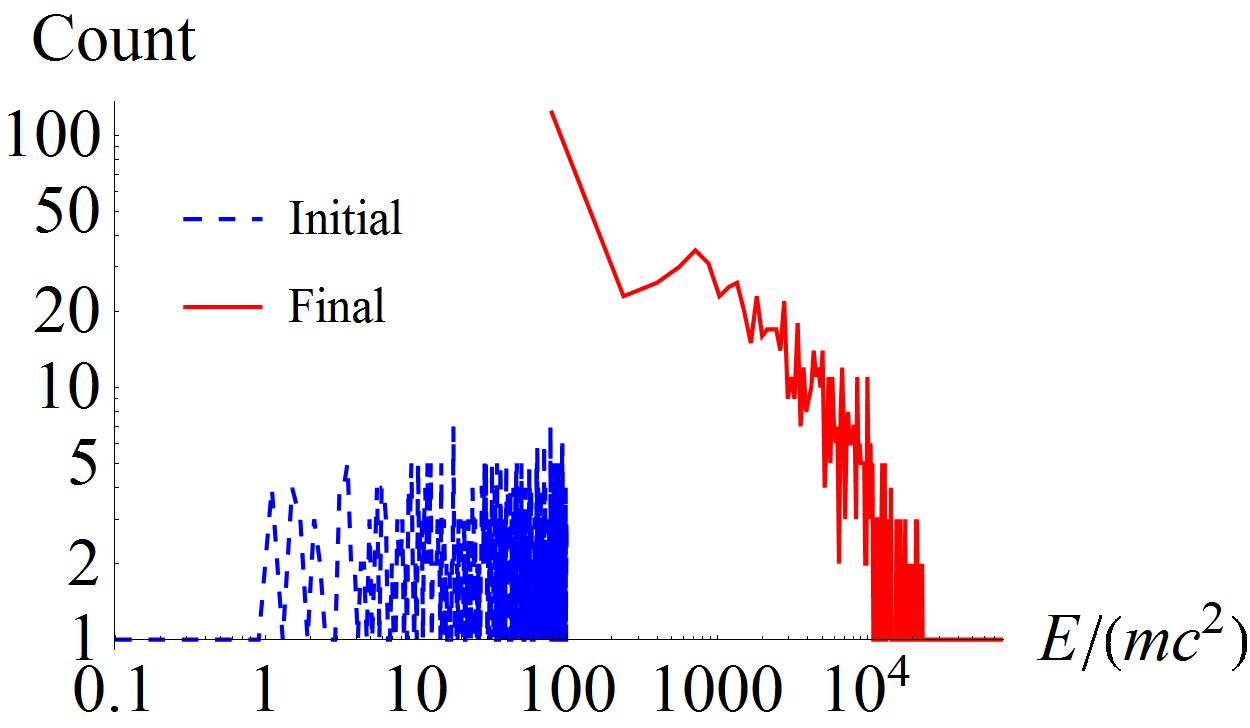}\label{fig:Configb}}  
\subfigure[]{\includegraphics[width=0.25\textwidth]{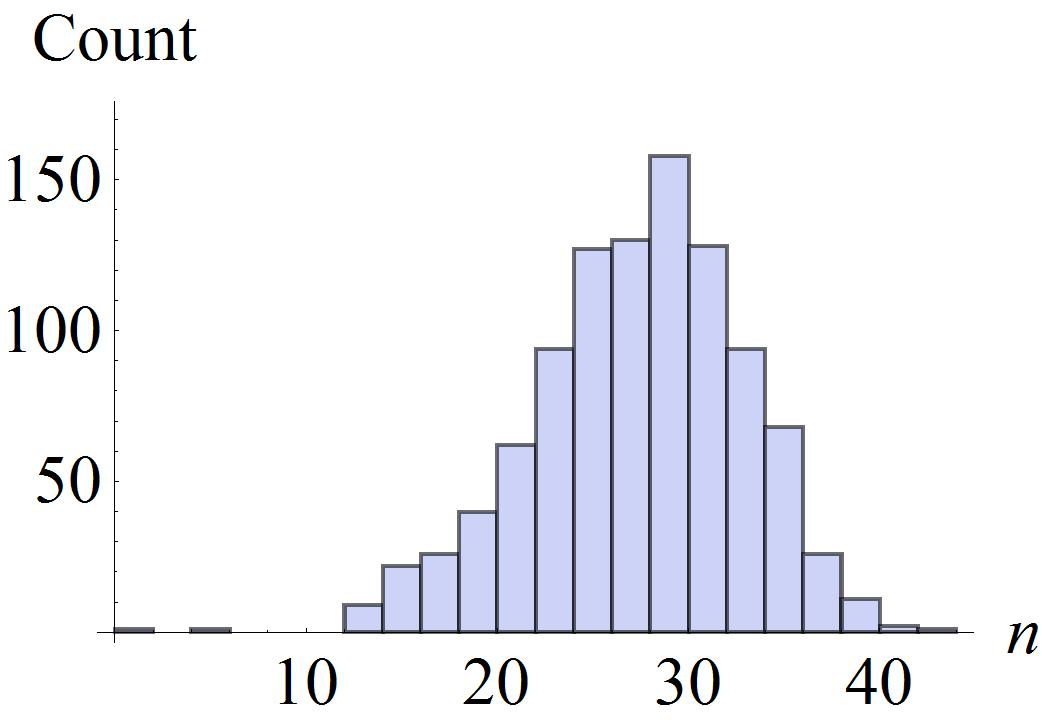}\label{fig:Configc}}\\
&\raisebox{0ex}[0ex][0ex]{\subfigure[]{\includegraphics[width=0.23\textwidth]{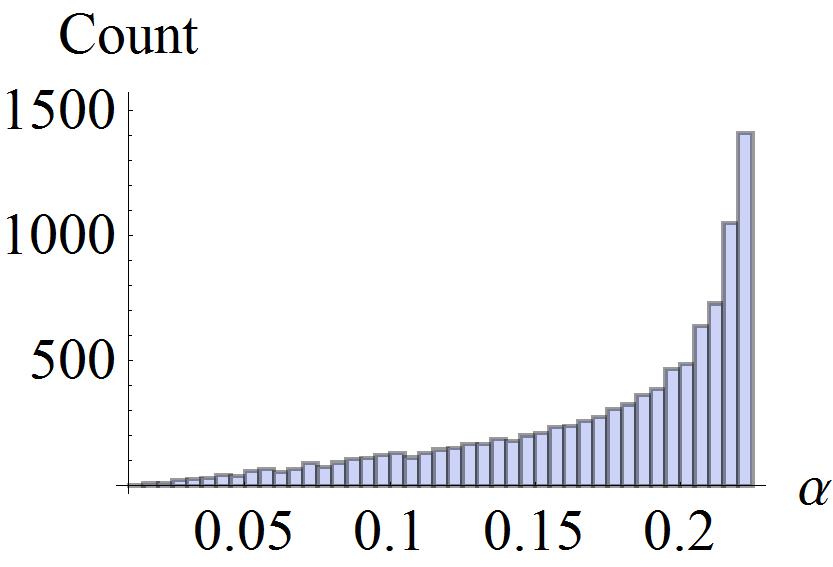}\label{fig:Configd}}}\qquad\qquad
\subfigure[]{\includegraphics[width=0.17\textwidth]{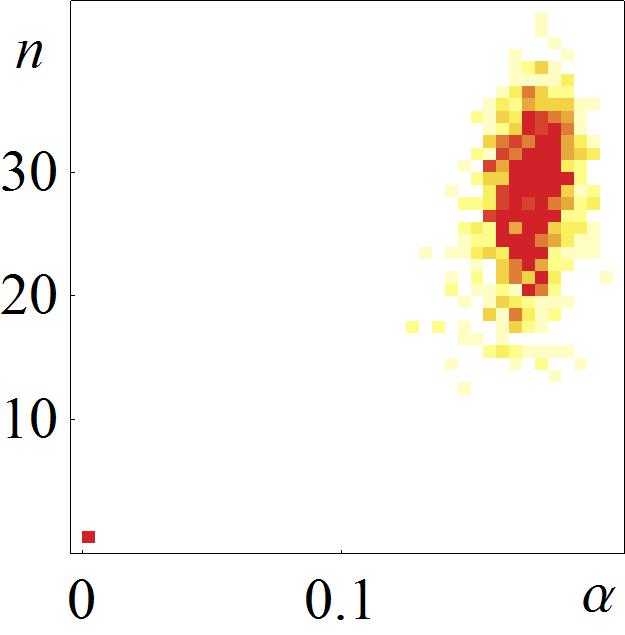}\label{fig:Confige}}
\end{tabular}
\caption{ (a) The layout of the system. The circle displays the shock front, the points stand for the particles and the short lines for the velocity vectors. (b) The energy spectrum for the initial (blue dashed) and final (red) states. The count stands for the number of particles at a given energy bin. (c) Histogram of the number of the nontrivial encounters $n$ where $\alpha\geqslant10^{-6}$. $\mu_{n}=26.9$ and $\sigma_{n}=5.5$. (d) Histogram of $\alpha$ in the mirror-reflection acceleration for 10000 particles. If $\beta=0.1$, then $\mu_{\alpha}=0.175$ and $\sigma_{\alpha}=0.048$. (e) The density pair-histogram shows a correlation between $n$ and $\alpha$. Their correlation is $R=0.874$.}
\end{figure*}

The total number of particles is 1000 (rather small so prone to make a great randomness in result) and that of time steps is 10000. The velocity of the shock is $v=0.1c$ (rather great in comparison with the typical SNR-supernova remnant), where $c$ stands for the speed of light and set to 3. The range of the whole square system is 500 (in a distance unit) and the radius of the circular shock front is 100. We show the configuration of the system in Fig.~\ref{fig:Configa}. We suppose the magnetic field to be 0.01 throughout the system. 

Figure~\ref{fig:Configb} shows the initial and final energy spectrum of particles. The initial spectrum is set to be a uniform distribution between $m c^2$ and $100m c^2$, where the mass $m$ of every particle is supposed 1. The tail-part of the final spectrum looks like a power-law distribution.

A geometrically growing system (which is called also  exponentially growing or multiplicative or proportionate system in other references) means the system where a representative quantity is growing in proportion to the current value of it as the system progresses. The system can be expressed by
\begin{linenomath} \begin{align}\label{eq:ansatz}
Z=(1+\alpha)^n Z_0,
\end{align} \end{linenomath}
where $Z$ is the representative quantity for the system, which is often called a size (e.g. the energy of a cosmic particle, population in a city or a mass in space), $\alpha$ is a growth rate, $n$ stands for the number of encounters of particles to the shock front (which is called the age later) and $Z_0$ is an initial value. $Z$ is apt to have a log-normal distribution according to the multiplicative central limit theorem, which says that the product of a large number of small random variables in normal distributions has to follow a log-normal distribution (\supno~\ref{note:lognormal}). The log-normal and the power-law are very similar but different in asymptotic behavior. 

In fact, in the above simulation, if $n$ is equal for every particle, then the final distribution of the particle energy should be log-normal. This makes us to inspect the distribution of $n$. We obtained a normal-like distribution for $n$ in this example (Fig.~\ref{fig:Configc}). Furthermore, its mean value is $\sim$30 that is much less than the total number of time steps (=10000). We inspected also the distribution of the growth rate $\alpha$ in the mirror-reflection acceleration (Fig.~\ref{fig:Configd}) where we took into account that what is uniformly distributed is not the incident angle but the impact parameter in flying to the circular shock front and the shock front is planar. The mean value of $\alpha$ is of order of $\beta=v/c$ because we adopted the first-order Fermi mode.

\section{General formalism of the power-law distribution for the geometrically growing system}\label{sec:formal}

In the above simulation the final spectrum looks like a power-law distribution at least at the tail part. Then which might be a principal factor for the final power-law distribution: the distribution of $\alpha$ or that of $n$? As we have seen in Fig.~\ref{fig:Configd}, the distribution of $\alpha$ appears nearly triangular with a skewness. A more important thing might be an approximate normal distribution of $n$. The diffusiveness of $n$ has not been expected before the simulation. 

Recall the general expression Eq.~\eqref{eq:ansatz} for the geometrically growing system. The logarithm of the final (or an instant) size $Z$ can be written as follows:
\begin{linenomath} \begin{align}
\label{eq:logZ} \ln Z&=n\ln(1+\alpha)+\ln Z_0  \\ 
\label{eq:logZapprox} &\approx n \alpha+\ln Z_0 , 
\end{align} \end{linenomath}
where we assume
\begin{linenomath} \begin{align} \label{eq:alphanegligible}
 \alpha\ll1
\end{align} \end{linenomath}
so $\ln(1+\alpha)\approx\alpha$. Note that $\alpha$ is a geometric mean of the growth rate for a particle because even for a given particle the growth rate could vary at every encounter. Provided that $\alpha\ll1$, this can be replaced with the arithmetic mean as shown in Fig.~\ref{fig:Configd}.

Here we can suppose that $n, \alpha$ and $\ln Z_0$ are normally distributed: the above simulation has shown it for $n$ and we extend this assumption to $\alpha$ and $\ln Z_0$ without loss of generality (\supno~\ref{note:normdistsup}). With this assumption, we can rewrite Eq.~\eqref{eq:logZapprox} in terms of a log-size $Y=\ln Z$:
\begin{linenomath} \begin{align}\label{eq:Y}
Y\approx n \alpha+Y_0,
\end{align} \end{linenomath}
where $Y_0=\ln Z_0$ is normally distributed by the above assumption. Hereafter we replace ``$\approx$'' with ``$=$.''

If we assume that $n, \alpha$ and $Y_0$ are normal-distributed variables, i.e.
\begin{linenomath} \begin{align}\label{eq:normalDist}
n\sim N(\mu_n, \sigma_n), \alpha\sim N(\mu_{\alpha}, \sigma_\alpha) \text{ and } Y_0\sim N(\mu_i, \sigma_i),
\end{align} \end{linenomath}
then they can be expressed in terms of random variables $x_n, x_{\alpha}$ and $x_i$ following the standard normal distribution $N(0,1)$, respectively:
\begin{linenomath} \begin{align}
n&=\mu_n+\sigma_n x_n,  \\ 
\alpha&=\mu_{\alpha}+\sigma_{\alpha} x_{\alpha}, \label{eq:normalDecomp}\\
Y_0&=\mu_i+\sigma_i x_i. \label{eq:normalY0}
\end{align} \end{linenomath}

If $x$s are completely independent, $Y$ in Eq.~\eqref{eq:Y} should be normally distributed and $Z$ be log-normal. Here we make another important assumption: $\alpha$ and $n$ might have a correlation (\supno~\ref{note:alphancor}). Figure~\ref{fig:Confige} in the above simulation shows that the correlation is $\sim$ 0.9. If the correlation is defined as 
\begin{linenomath} \begin{align}\label{eq:R}
R=\texttt{cor}(n, \alpha)=\texttt{cor}(x_n, x_{\alpha}), 
\end{align} \end{linenomath}
where $-1<R<1$, then $x_n$ and $x_{\alpha}$ can be modeled as (\supno~\ref{note:orthsum})
\begin{linenomath} \begin{align}\label{eq:xcor}
x_n=\sqrt{\vert R \vert}x_c+\sqrt{1-\vert R \vert}x_{nnc}, \\
x_{\alpha}=\sqrt{\vert R \vert}x_c+\sqrt{1-\vert R \vert}x_{\alpha nc} \notag
\end{align} \end{linenomath}
and $Y$ can be rewritten as
\begin{linenomath} \begin{align}\label{eq:Y1}
Y_1=(\mu_n+\sigma_n\sqrt{\vert R \vert}x_c)(\mu_{\alpha}+\texttt{sgn}(R)\sigma_{\alpha}\sqrt{\vert R \vert}x_c)+(\mu_i+\sigma_{i1}x_{i1}),
\end{align} \end{linenomath}
where $x_c, x_{nnc}, x_{\alpha nc}$ and $x_{i1}$ are all i.i.d (independently and identically distributed) random variables with $N(0,1)$ and $\texttt{sgn}(R)$ stands for the sign of $R$. This sign can be multiplied to either $\sigma_n$ or $\sigma_{\alpha}$ alternatively in the above equation. And $\sigma _{i1}$ is determined as in Eq.~\eqref{eq:d}.

We manipulate Eq.~\eqref{eq:Y1} to ``complete the square'':
\begin{linenomath} \begin{align}\label{eq:Y2}
Y_2=\texttt{sgn}(R)(a x_c+b)^2 +c+d x_{i1},
\end{align} \end{linenomath}
where
\begin{linenomath} \begin{align}
a&=\sqrt{\sigma_{\alpha} \sigma_n \vert R \vert}, \label{eq:a} \\ 
b&=\dfrac{\vert \mu_{\alpha} \sigma_n + \texttt{sgn}(R)\mu_n \sigma_{\alpha} \vert}{2\sqrt{\sigma_{\alpha} \sigma_n}}, \label{eq:b}\\
c&=\mu_i+\mu_n \mu_{\alpha}-\texttt{sgn}(R)b^2, \label{eq:c} \\
d&=\sigma_{i1}=\sqrt{\sigma_i^2+(\mu_n^2\sigma_{\alpha}^2+\mu_{\alpha}^2\sigma_n^2)(1-\vert R \vert)+\sigma_n^2 \sigma_{\alpha}^2(1-R^2)},\label{eq:d}
\end{align} \end{linenomath}
The norm in $b$ comes from that $\texttt{sgn}(R)$ in Eq.~\eqref{eq:Y1} should belong alternatively to either $n$ or $\alpha$. As we can see, when the system evolves, i.e. the time goes by, the parameters $b$, $c$ and $d$ vary mainly due to $\mu_n$. So the distribution of $Y_2$ will vary and our analysis will be dynamic.

We can decompose $Y_2$ in Eq.~\ref{eq:Y2} into two terms: $Y_{21}=\texttt{sgn}(R)(a x_c+b)^2$ which follows a chi-square-like distribution and $Y_{22}=c+d x_{i1}$ following the normal distribution. The probability density function (PDF) of $Y_{21}$ will be discussed later in detail and can be expressed by Eq.~\eqref{eq:Y3PDF}. The PDF of $Y_2=Y_{21}+Y_{22}$ can be calculated by integrals $p_{Y_2}(Y)=\int^{\infty}_{-\infty}p_{Y_{21}}(Y-u)p_{Y_{22}}(u)du$ or $p_{Y_2}(Y)=\int^{\infty}_{-\infty}p_{Y_{21}}(u)p_{Y_{22}}(Y-u)du$ \citep[e.g., see][or any references on the probability theory]{Gwang-so2009}, where we regard that $Y_{21}$ and $Y_{22}$ are independent. The PDFs of $Y_{21}, Y_{22}$ and $Y_2$ can be written as follows:
\begin{linenomath} \begin{align}
p_{Y_{21}}(Y)&=\frac{\exp\left[-\frac{1}{2} \left(\frac{\sqrt{\texttt{sgn}(R)Y}-b}{a} \right)^2 \right]}{2\sqrt{2\pi}a \sqrt{\texttt{sgn}(R)Y}},\label{eq:Y21PDF}\\
p_{Y_{22}}(Y)&=\frac{1}{\sqrt{2\pi}d}\exp\left[-\dfrac{\left(Y-c\right)^2}{2d^2}\right],\label{eq:Y22PDF}\\ 
p_{Y_2}(Y)&=\int\limits^{\footnotesize \begin{subarray}{1} u=Y \text{ in } R>0 \\  u=+\infty \text{ in } R\leqslant0 \end{subarray}}
_{\footnotesize \begin{subarray}{1} u=-\infty \text{ in } R>0 \\  u=Y \text{ in } R\leqslant0 \end{subarray}}
\frac{\exp\left[-\frac{\left(\sqrt{\texttt{sgn}(R)(Y-u)}-b\right)^2}{2a^2}-\frac{\left(u-c\right)^2}{2d^2}\right]}{4\pi a d\sqrt{\texttt{sgn}(R)(Y-u)}}du \label{eq:Y2PDF1}\\
&=\int\limits^{\footnotesize \begin{subarray}{1} u=+\infty \text{ in } R>0 \\  u=0 \text{ in } R\leqslant0 \end{subarray}}
_{\footnotesize \begin{subarray}{1} u=0 \text{ in } R>0 \\  u=-\infty \text{ in } R\leqslant0 \end{subarray}}
\frac{\exp\left[-\frac{\left(\sqrt{\texttt{sgn}(R)u}-b\right)^2}{2a^2}-\frac{\left(Y-u-c\right)^2}{2d^2}\right]}{4\pi a d\sqrt{\texttt{sgn}(R)u}}du \label{eq:Y2PDF2}
\end{align} \end{linenomath}
Integrals Eq.~\eqref{eq:Y2PDF1} and ~\eqref{eq:Y2PDF2} are interchangeable through a variable transformation $u\leftrightarrow Y-u$ and seem to have no analytic solution. As we will see later, the case $R=0$ belongs to the log-normal approximation so should be attached to cases $R<0$ if we consider the slope only at the upper limit.

We can obtain a distribution of $Z$ in Eq.~\eqref{eq:ansatz} \citep[e.g.][]{Gwang-so2009}:
\begin{linenomath} \begin{align}
Z_2&=\exp{Y_2}, \label{eq:YtoZ}\\
p_{Z_2}(Z)=\left\vert \tfrac{dY}{dZ} \right\vert p_{Y_2}(Y)&=\tfrac{1}{Z}p_{Y_2}(Y)=e^{-Y}p_{Y_2}(Y). \label{eq:Z2PDF}
\end{align} \end{linenomath}
However, analytic solution of $p_{Z_2}(Z)$ looks impossible due to $p_{Y_2}(Y)$.

We will call $Y_2$ in Eq.~\eqref{eq:Y2}, $p_{Y_2}(Y)$ in Eq.~\eqref{eq:Y2PDF1},~\eqref{eq:Y2PDF2} and  $p_{Z_2}(Z)$ in Eq.~\eqref{eq:Z2PDF} ``correct forms'' in comparison with the approximations to themselves mentioned below. The meaning of ``correct'' is relative since $Y_2$ itself is an approximation to the logarithm of the geometrically growing system Eq.~\eqref{eq:ansatz}. We can formulate a proposition.

\begin{proposition} \label{th:prop1}
Assuming Eq.~\eqref{eq:alphanegligible}, Eq.~\eqref{eq:normalDist} and Eq.~\eqref{eq:R}, the logarithm of a representative quantity (or size) of the geometrically growing system Eq.~\eqref{eq:ansatz} can be approximate by Eq.~\eqref{eq:Y2}.
\end{proposition}

The later simulations show that these ``correct forms'' approximate the practical distribution of raw data very well. Hereafter, we identify ``the correct forms'' with the geometrically growing system Eq.~\eqref{eq:ansatz} itself as far as Eq.~\eqref{eq:alphanegligible}, Eq.~\eqref{eq:normalDist} and Eq.~\eqref{eq:R} are assumed. 

Since Eq.~\eqref{eq:Y2PDF1} and ~\eqref{eq:Y2PDF2} have no analytic solution, we will make approximations to that. First, we can take an approximation of type of the above $Y_{21}$ which seems to follow a chi-square-like distribution. On this way $x_{i1}$ can be embraced into the first-order term of $x_c$ in the first quadratic term in Eq.~\eqref{eq:Y2}: 
\begin{linenomath} \begin{align}
Y_2&=\texttt{sgn}(R)(a^2 x_c^2+2ab x_c+b^2) +c+d x_{i1} \\
&=\texttt{sgn}(R)a^2 x_c^2+\sqrt{4a^2b^2+d^2}x_3+\texttt{sgn}(R)b^2+c \label{eq:Y2toY3}\\
&=\texttt{sgn}(R)a^2 x_3^2+\sqrt{4a^2b^2+d^2}x_3+\texttt{sgn}(R)b^2+c.
\end{align} \end{linenomath}
In the second equation we consider that $x_c$ and $x_{i1}$ are independent and they compose a single normal random variable $x_3$ which differs from the previous $x$s but follows also $N(0,1)$ distribution (\supno~\ref{note:orthsum}). In the last equation $x_c^2$ is replaced with $x_3^2$ by using $\texttt{cor}(x_c^2, x_c)=0$, $\texttt{cor}(x_c^2, x_3)=0$ and $\texttt{cor}(x_3^2, x_3)=0$.
Thus we can transform $Y_2$ to a completely squared form of a single random variable $x_3$:
\begin{linenomath} \begin{align}\label{eq:Y3}
Y_3=\texttt{sgn}(R)(A x_3+B)^2 +C,
\end{align} \end{linenomath}
where
\begin{linenomath} \begin{align}\label{eq:ABC}
A&=a, \notag \\ 
B&=\dfrac{\sqrt{4a^2b^2+d^2}}{2a}, \\
C&=c+\texttt{sgn}(R)b^2-\texttt{sgn}(R)B^2=c-\texttt{sgn}(R)\frac{d^2}{4a^2}. \nonumber 
\end{align} \end{linenomath}

We call the distribution of $Y_3$ in Eq.~\eqref{eq:Y3} a completely squared chi ($\chi$) distribution with 1 degree of freedom (shortly, CS$\chi_1$). This distribution can be derived through transformation of the chi square distribution with 1 degree of freedom ($\chi_1^2$) of $X$ where $X=x_3^2$ can be rewritten as
\begin{linenomath} \begin{align}\label{eq:XfromY}
X=\left(\frac{\sqrt{\texttt{sgn}(R)(Y-C)}-B}{A} \right)^2.
\end{align} \end{linenomath}
If we take into account of Eq.~\eqref{eq:Z2PDF} and the PDF of $\chi_1^2$ distribution $\chi_1^2(X)=\tfrac{1}{\sqrt{2\pi X}}\exp\left( -X/2 \right)$, we can obtain the PDF of $Y_3$, which is just the CS$\chi_1$ distribution:
\begin{linenomath} \begin{align}\label{eq:Y3PDF}
p_{Y_3}(Y)=\left\vert \frac{\sqrt{\texttt{sgn}(R)(Y-C)}-B}{A^2\sqrt{\texttt{sgn}(R)(Y-C)}} \right\vert \chi_1^2\left[ \left(\frac{\sqrt{\texttt{sgn}(R)(Y-C)}-B}{A} \right)^2 \right]=\frac{\exp\left[-\dfrac{1}{2} \left(\frac{\sqrt{\texttt{sgn}(R)(Y-C)}-B}{A} \right)^2 \right]}{2\sqrt{2\pi}A \sqrt{\texttt{sgn}(R)(Y-C)}},
\end{align} \end{linenomath}
where a normalization factor 1/2 is multiplied. The distribution of $Z_3=e^{Y_3}$ is obtained analytically via Eq.~\eqref{eq:Z2PDF}
\begin{linenomath} \begin{align}\label{eq:Z3PDF}
p_{Z_3}(Z)=\frac{\exp\left[-\dfrac{1}{2} \left(\frac{\sqrt{\texttt{sgn}(R)(\ln Z-C)}-B}{A} \right)^2 \right]}{2\sqrt{2\pi}A Z\sqrt{\texttt{sgn}(R)(\ln Z-C)}},
\end{align} \end{linenomath}
The domain of $Z$ is $(e^{C}, +\infty)$ for $R> 0$ and $(0,e^{C})$ for $R\leqslant0$. We should call this distribution Eq.~\eqref{eq:Z3PDF} a log-completely squared chi ($\chi$) distribution with 1 degree of freedom (shortly, log-CS$\chi_1$). 

We can make an alternative approximation of $Y_2$ in Eq.~\eqref{eq:Y2} in the form of normal distribution such as $Y_{22}$. In Eq.~\eqref{eq:Y2toY3}, if we regard $x_c^2$ as a random variable with the mean of 1 and the variance of 2 (in fact, $x_c^2$ follows the chi-square distribution) and use $\texttt{cor}(x_c^2, x_{i2})=0$, we can obtain another, normal-distributed, approximation to $Y_2$:
\begin{linenomath} \begin{align}\label{eq:Y4}
Y_4=F+G x_4,
\end{align} \end{linenomath}
where $x_4$ is a standard normal random variable different from the above $x$s and 
\begin{linenomath} \begin{align}\label{eq:FG}
F&=\texttt{sgn}(R)(a^2+b^2)+c, \notag \\ 
G&=\sqrt{2a^4+4a^2b^2+d^2}.  
\end{align} \end{linenomath}
The PDFs of $Y_4$ and $Z_4=e^{Y_4}$ are evaluated normal and log-normal via Eq.~\eqref{eq:Z2PDF}:
\begin{linenomath} \begin{align}
p_{Y_4}(Y)&=\frac{1}{\sqrt{2\pi}G}\exp\left[-\dfrac{\left(Y-F\right)^2}{2G^2}\right],\label{eq:Y4PDF}\\ 
p_{Z_4}(Z)&=\frac{1}{\sqrt{2\pi}G Z}\exp\left[-\dfrac{\left(\ln Z-F\right)^2}{2G^2}\right]. \label{eq:Z4PDF}
\end{align} \end{linenomath}

We can check that the above $Y_1, Y_2, Y_3$ and $Y_4$ have the same mean of $F$ and variance of $G$ by assuming that the above $x$s and their squares are independent and even normally distributed. Then, which among the corresponding ``analytic approximations'' $p_{Z_3}(Z)$ and $p_{Z_4}(Z)$ is closer to ``the correct form'' $p_{Z_2}(Z)$? To answer this question, we analyze the profile of the $p_Z(Z)$s in log-log diagram, especially their asymptotic behavior. 

\begin{figure*}
\centering
\subfigure[]{\includegraphics[width=0.3\textwidth]{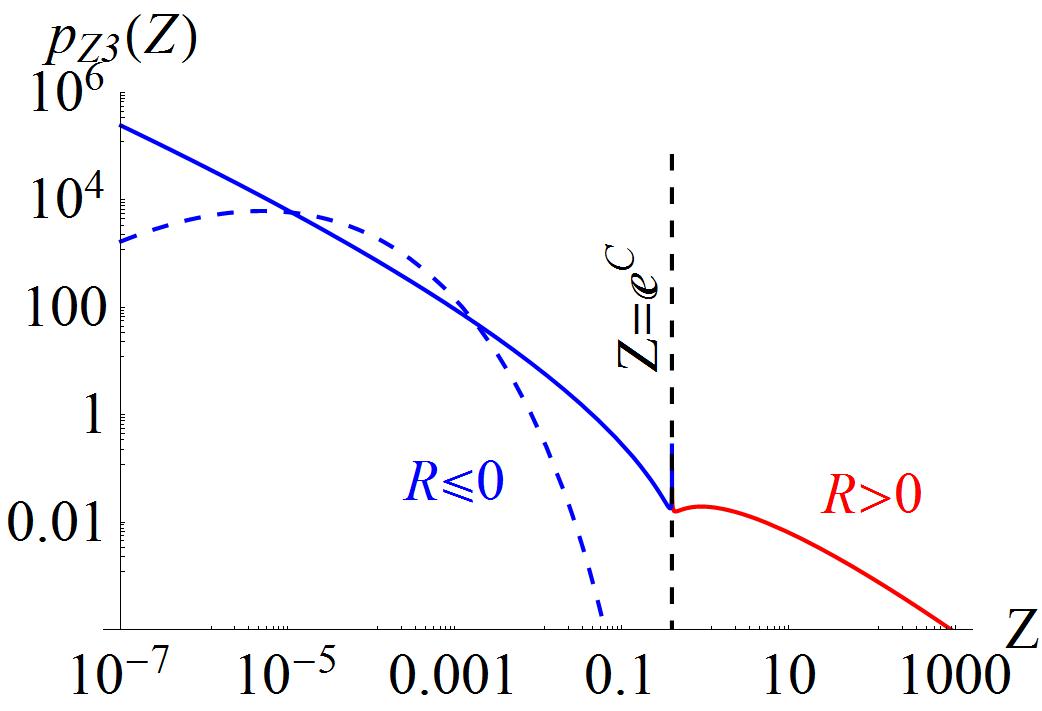}\label{fig:logCSC1a}}  
\subfigure[]{\includegraphics[width=0.27\textwidth]{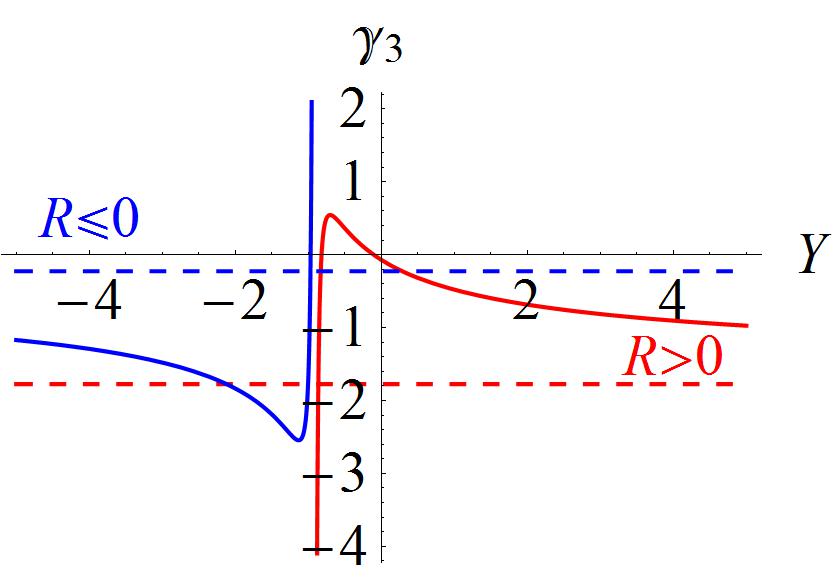}\label{fig:logCSC1c}}  \quad 
\subfigure[]{\includegraphics[width=0.25\textwidth]{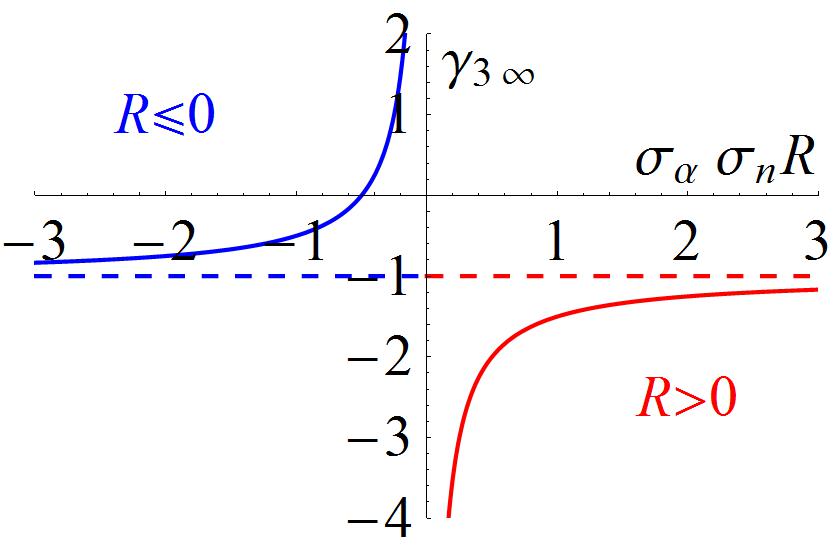}\label{fig:logCSC1b}}
\caption{\label{fig:logCSC1} (a) The profile of the log-CS$\chi_1$ distribution with $R>0$ (red) and $R\leqslant0\ \&\ \gamma_{3\infty}<0$ (blue) and $R\leqslant0\ \&\ \gamma_{3\infty}\geqslant0$ (blue dashed). The vertical dashed line stands for the singularity at $Z=e^{C}$. (b) Variation of the slope $\gamma$ vs. $Y$ for the log-CS$\chi_1$.  The dashed lines stand for $\gamma_{3\infty}$ with $R>0$ (red) and $R\leqslant0$ (blue) respectively. (c) The asymptotic slope $\gamma_{3\infty}$ vs. the product $\sigma_{\alpha}\sigma_n R$ for the log-CS$\chi_1$. The dashed line stands for $\gamma_{\infty}=-1$.} 
\end{figure*}

First, let consider $p_{Z_3}(Z)$ in Eq.~\eqref{eq:Z3PDF}. The $p_{Z_3}(Z)$ has different shapes depending on the sign of $R$ (Fig.~\ref{fig:logCSC1a}). There appears a singularity at $Z=e^{C}$, which becomes a lower or upper limit of the distribution with various $\texttt{sgn}(R)$. The exponent of the power-law or the slope in log-log diagram $\gamma$ can be calculated by 
\begin{linenomath} \begin{align}\label{eq:gamma3}
\gamma_3&=\frac{d(\ln(p_{Z_3}(Z)))}{d(\ln Z)}=\frac{d\left(-Y-\ln (\sqrt{2\pi}A)-\frac{1}{2}\ln (\texttt{sgn}(R)(Y-C))-\frac{1}{2} \left(\frac{\sqrt{\texttt{sgn}(R)(Y-C)}-B}{A} \right)^2 \right)}{dY} \nonumber \\
&=-1-\dfrac{1}{2(Y-C)}-\texttt{sgn}(R)\dfrac{\sqrt{\texttt{sgn}(R)(Y-C)}-B}{2A^2\sqrt{\texttt{sgn}(R)(Y-C)}}, 
\end{align} \end{linenomath}
where we substitute $Z=e^Y$ into Eq.~\eqref{eq:Z3PDF}. A variation of $\gamma_3$ vs. $Y$ is exemplified in Fig.~\ref{fig:logCSC1c}. The slope converges to a constant at the limit of $Y\rightarrow +\infty$ ($Z\rightarrow+\infty$) for $R>0$ or $Y\rightarrow -\infty$ ($Z\rightarrow 0$) for $R\leqslant0$:
\begin{linenomath} \begin{align}\label{eq:gamma3lim}
\gamma_{3\infty}=\lim_{\footnotesize \begin{subarray}{1} Y\to +\infty (Z\to +\infty) \text{ in } R>0 \\  Y\to-\infty(Z\to 0) \text{ in } R\leqslant0 \end{subarray} } \gamma_3=-\left(1+\texttt{sgn}(R)\dfrac{1}{2A^2}\right)=-\left(1+\dfrac{1}{2\sigma_{\alpha}\sigma_n R}\right).
\end{align} \end{linenomath}
So we obtain an asymptotic power-law behavior. On the opposite sides of $Z$ we can expect the aforementioned singularity. The asymptotic slope $\gamma_{\infty}$ depends only on $A$, i.e. product of $\sigma_{\alpha}$, $\sigma_n$ and $R$ (Fig.~\ref{fig:logCSC1b}) but not on the initial distribution of $Z_0$ or any mean values. Changing $A$, we can obtain the slope $\gamma_{\infty}$ from $-\infty$ to $+\infty$, which can explain the slope of any power-laws. If the variances of $\alpha$ and $n$ are smaller, we could obtain a greater slope. If $R=0$, then an infinitive $\gamma_{\infty}$ is given, similarly to the log-normal distribution. Even if $\gamma_{\infty}$ could be finite, we can not draw an asymptotic line in a finite region in log-log diagram. In other words, if we write as 
\begin{linenomath} \begin{align}
\lim_{Z\to\infty}p_Z(Z)=C'Z^{\gamma_\infty}\label{eq:pzpower},
\end{align} \end{linenomath}
where $C'$ is a constant and $R$ is assumed positive, we can find that $C'$ diverges at that limit. The profile has two stationary points for $R>0$ and one or two depending on $\gamma_{\infty}$ for $R\leqslant0$ (Fig.~\ref{fig:logCSC1a}). If $R\leqslant0$ and $\gamma_{\infty}<0$, then the profile with  looks like a power-law with a cutoff. But, if $R\leqslant0$ and $\gamma_{\infty}\geqslant 0$, the profile lowers at the limit of $Z\rightarrow 0$ like the blue dashed curve shown in Fig.~\ref{fig:logCSC1a}. We can find a convex low part corresponding to the local maximum in the cases of $R>0$ and $R\leqslant0\ \&\ \gamma_{3\infty}\geqslant0$. This could explain a kind of ``roll-over'' in various phenomena. But this does not appear in the case of $R\leqslant0\ \&\ \gamma_{3\infty}<0$.

\begin{lemma} \label{th:lem1}
A quantity whose logarithm is expressed as Eq.~\eqref{eq:Y3} has a log-CS$\chi_1$ distribution Eq.~\eqref{eq:Z3PDF}. This distribution has an asymptotic power-law behavior. The asymptotic exponent of the power-law or the asymptotic slope in log-log diagram, which is determined by Eq.~\eqref{eq:gamma3lim}, is only related to the inverse of the product of the standard deviations of the system parameters and their mutual correlation but independent of an initial distribution of the system or any mean value of parameters.
\end{lemma}

We can say that if the system Eq.~\eqref{eq:ansatz} is expressed by $Z_3$, the asymptotic slope is inversely proportional to the product of the standard deviations of the growth rate and the age and the correlation between them but independent of the mean values of them or an initial distribution of the system.

Next, let consider $p_{Z_4}(Z)$ in Eq.~\eqref{eq:Z4PDF}. This distribution is log-normal and its property has been widely studied. This has no singularity like $p_{Z_3}(Z)$. Instead, its asymptotic slope diverges on both sides:
\begin{linenomath} \begin{align}
\gamma_4&=\dfrac{d(\ln(p_{Z_4}(Z)))}{d(\ln Z)}=\dfrac{d(-Y-\ln (\sqrt{2\pi}G)-\frac{(Y-F)^2}{2G^2})}{dY}=-1-\frac{Y-F}{G^2}, \label{eq:gamma4}\\ 
\gamma_{4\infty}&=\lim_{Y\to \pm\infty} \gamma_4=\mp\infty,\label{eq:gamma4lim}
\end{align} \end{linenomath}
where the limit $Y\rightarrow +\infty$ ($Y\rightarrow -\infty$) corresponds to $Z\rightarrow +\infty$ ($Z\rightarrow 0$).

Now we can ask what the asymptotic slope of ``the correct form'' itself $p_{Z_2}(Z)$ in Eq.~\eqref{eq:Z2PDF} is. In order to clarify this, we consider first the asymptotic behavior of the general form of $p_Y(Y)$. Use Eq.~\eqref{eq:YtoZ},~\eqref{eq:Z2PDF} and Eq.~\eqref{eq:pzpower}, deriving
\begin{linenomath} \begin{align}
p_Y(Y)=e^Y p_Z(Z)=e^Y C'(e^Y)^{\gamma_\infty}=C'e^{Y(\gamma_\infty+1)}.
\end{align} \end{linenomath}
Then we can obtain
\begin{linenomath} \begin{align}
\frac{d(\ln p_Y(Y))}{dY}=\gamma_\infty+1. \label{eq:lnpyslope}
\end{align} \end{linenomath}
In the case of the log-CS$\chi_1$, we obtain from Eq.~\eqref{eq:gamma3lim}
\begin{linenomath} \begin{align}
\gamma_{3\infty}+1=-\texttt{sgn}(R)\dfrac{1}{2A^2}=-\texttt{sgn}(R)\dfrac{1}{2a^2}=-\dfrac{1}{2\sigma_{\alpha}\sigma_n R},
\end{align} \end{linenomath} 
where $C'$ also diverges.

\begin{figure*}
\centering
\subfigure[]{\includegraphics[width=0.24\textwidth]{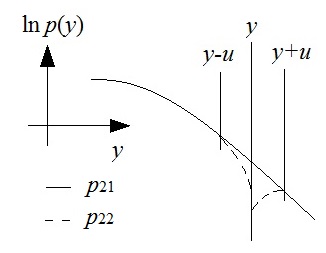}}
\subfigure[]{\includegraphics[width=0.24\textwidth]{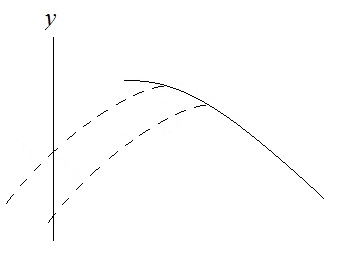}}
\subfigure[]{\includegraphics[width=0.24\textwidth]{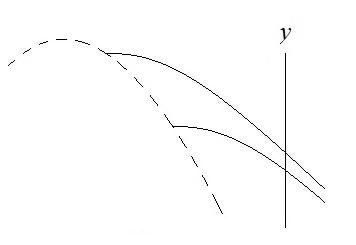}}
\subfigure[]{\includegraphics[width=0.24\textwidth]{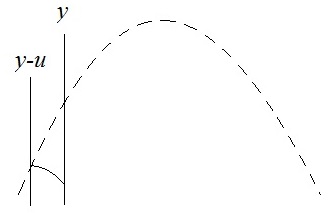}}
\caption{\label{fig:convolution} The convolution of $p_{Y_{21}}$ in Eq.~\eqref{eq:Y21PDF} and $p_{Y_{22}}$ in Eq.~\eqref{eq:Y22PDF} in the case $R>0$. (a) and (b) show the PDFs in the positive and negative infinities, respectively, for Eq.~\eqref{eq:Y21PDF} and (c) and (d) show the same for Eq.~\eqref{eq:Y22PDF}. The ordinate stands for the logarithm of PDF and the abscissa for $Y$ since the logarithm of the PDFs could show asymptotic behaviors from Eq.~\eqref{eq:lnpyslope}. The vertical lines imply only the places for $y$, $y-u$ and $y+u$.} 
\end{figure*}

Let inspect the asymptotic behavior of $p_{Y_2}(Y)$. We consider first the case of $R> 0$. We have obtained two kinds of expressions for $p_{Y_2}(Y)$ in Eq.~\eqref{eq:Y2PDF1} and Eq.~\eqref{eq:Y2PDF2}. First, we use Eq.~\eqref{eq:Y2PDF1} to induce the asymptotic behavior. The integrand can be imagined as in Fig.~\ref{fig:convolution}(a) and ~\ref{fig:convolution}(b) corresponding to the cases of $Y\to+\infty$ and $Y\to-\infty$, respectively. Note that the integrand has a form of convolution between $p_{Y_{21}}(Y)$ (Eq.~\ref{eq:Y21PDF}) and $p_{Y_{22}}(Y)$ (Eq.~\ref{eq:Y22PDF}). We recall that the $p_{Y_{22}}(Y)$, which is the normal distribution, has steeper slope than $p_{Y_{21}}(Y)$, which is the CS$\chi_1$ distribution, at the limit $Y\to+\infty$. Therefore the asymptotic behavior is determined mainly by $p_{Y_{21}}(Y)$ at that limit. Meanwhile, this depends mainly on $p_{Y_{22}}(Y)$ at the limit $Y\to-\infty$ because $p_{Y_{21}}(Y)$ has a semi-infinite domain of $Y$ while $p_{Y_{22}}(Y)$ has a bi-infinite and symmetric domain. This means that the dependency on $Y$ should be attributed to $p_{Y_{21}}(Y)$ at the limit $Y\to+\infty$ but to $p_{Y_{22}}(Y)$ at the limit $Y\to-\infty$ in spite of that $Y$ is included in $p_{Y_{21}}(Y)$ in Eq.~\eqref{eq:Y2PDF1}. Similarly, in the form of Eq.~\eqref{eq:Y2PDF2} the dependency on $Y$ should be attributed to $p_{Y_{21}}(Y)$ at the limit $Y\to+\infty$ but $p_{Y_{22}}(Y)$ at the limit $Y\to-\infty$. Namely, the both expressions Eq.~\eqref{eq:Y2PDF1} and Eq.~\eqref{eq:Y2PDF2} of $p_{Y_2}(Y)$ reach to the same formats at the limits:
\begin{linenomath} \begin{align}
\lim_{Y\to+\infty}p_{Y_2}(Y)&=
\int\limits^{Y}_{-\infty}\frac{\exp\left[-\frac{\left(\sqrt{Y-u}-b\right)^2}{2a^2}-\frac{\left(u-c\right)^2}{2d^2}\right]}{4\pi a d\sqrt{Y-u}}du, \label{eq:Y2PDF11}\\
\lim_{Y\to-\infty}p_{Y_2}(Y)&=\int\limits^{+\infty}_{0}
\frac{\exp\left[-\frac{\left(\sqrt{u}-b\right)^2}{2a^2}-\frac{\left(Y-u-c\right)^2}{2d^2}\right]}{4\pi a d\sqrt{u}}du. \label{eq:Y2PDF21}
\end{align} \end{linenomath}

In Fig.~\ref{fig:convolution}(a) we can see that $\exp\left[-\frac{\left(u-c\right)^2}{2d^2}\right]$ drops so steeply that only a finite domain of $u$ around $c$ can give a nontrivial contribution to the integral. So we can regard that $u$ is much smaller than $Y$ in Eq.~\eqref{eq:Y2PDF11}. We can also assume that outside a finite domain of $u$ the term $\exp\left[-\frac{\left(\sqrt{u}-b\right)^2}{2a^2}\right]$ could be negligible. Therefore we can replace $Y-u$ with $Y$ and the integrals over a semi-infinite domain of $u$ with an integral over the bi-infinite domain of $u$, which can make the above integrals much simpler: the both parts of $p_{Y_{21}}(Y)$ and $p_{Y_{22}}(Y)$ can be perfectly divided. We can extend the same discussion to the case of $R\leqslant0$. At last we obtain the asymptotic distributions of ``the correct form'' PDF $p_{Y_2}(Y)$ as
\begin{linenomath} \begin{align}
p_{Y_{31}}(Y)=\lim_{\texttt{sgn}(R)Y\to+\infty}p_{Y_2}(Y)&=\frac{\exp\left[-\frac{\left(\sqrt{\texttt{sgn}(R)Y}-b\right)^2}{2a^2}\right]}{2\sqrt{2\pi} a \sqrt{\texttt{sgn}(R)Y}}
\int\limits^{+\infty}_{-\infty}\frac{\exp\left[-\frac{\left(u-c\right)^2}{2d^2}\right]}{\sqrt{2\pi} d}du
=\frac{\exp\left[-\frac{\left(\sqrt{\texttt{sgn}(R)Y}-b\right)^2}{2a^2}\right]}{2\sqrt{2\pi} a \sqrt{\texttt{sgn}(R)Y}}, \label{eq:Y31PDF}\\
p_{Y_{41}}(Y)=\lim_{\texttt{sgn}(R)Y\to-\infty}p_{Y_2}(Y)&=\frac{\exp\left[\frac{\left(Y-c\right)^2}{2d^2}\right]}{\sqrt{2\pi} d}
\int\limits^{+\infty}_{-\infty}\frac{\exp\left[-\frac{\left(\sqrt{\texttt{sgn}(R)u}-b\right)^2}{2a^2}\right]}{2\sqrt{2\pi} a\sqrt{\texttt{sgn}(R)u}}du
=\frac{\exp\left[\frac{\left(Y-c\right)^2}{2d^2}\right]}{\sqrt{2\pi} d}, \label{eq:Y41PDF}
\end{align} \end{linenomath}
which are the PDF of 
\begin{linenomath} \begin{align}
Y_{31}&=\texttt{sgn}(R)(a x_{31}+b)^2, \label{eq:Y31}\\
Y_{41}&=c+d x_{41}, \label{eq:Y41}
\end{align} \end{linenomath}
respectively, where $x_{31}$ and $x_{41}$ are $N(0,1)$ random variable. The corresponding $p_{Z}(Z)$s are
\begin{linenomath} \begin{align}
p_{Z_{31}}(Z)=\lim_{\texttt{sgn}(R)Y\to+\infty}p_{Z_2}(Z)&
=\frac{1}{2\sqrt{2\pi} a Z\sqrt{\texttt{sgn}(R)\ln Z}}\exp\left[-\frac{\left(\sqrt{\texttt{sgn}(R)\ln Z}-b\right)^2}{2a^2}\right] \label{eq:Z31PDF}\\
p_{Z_{41}}(Z)=\lim_{\texttt{sgn}(R)Y\to-\infty}p_{Z_2}(Z)&
=\frac{1}{\sqrt{2\pi} d Z} \exp\left[\frac{\left(\ln Z-c\right)^2}{2d^2}\right]. \label{eq:Z41PDF}
\end{align} \end{linenomath}
Those are very similar to Eq.~\eqref{eq:Z3PDF} and ~\eqref{eq:Z4PDF}. Here the limit $\texttt{sgn}(R)Y\to+\infty$ corresponds to $Z\to +\infty \text{ for } R>0$ and $Z\to 0 \text{ for } R\leqslant0$ while $\texttt{sgn}(R)Y\to-\infty$ to $Z\to 0 \text{ for } R>0$ and $Z\to +\infty \text{ for } R\leqslant0$.

Thus we can be sure that $p_{Z_2}(Z)$ in Eq.~\eqref{eq:Z2PDF} has the same asymptotic behaviors as both the log-CS$\chi_1$ and log-normal distributions in Eq.~\eqref{eq:gamma3lim} and ~\eqref{eq:gamma4lim}:
\begin{linenomath} \begin{align}\label{eq:gamma2lim}
\gamma_{2\infty}&=\lim_{\footnotesize \begin{subarray}{1} Y\to +\infty (Z\to +\infty) \text{ in } R>0 \\  Y\to-\infty(Z\to 0) \text{ in } R\leqslant0 \end{subarray}  } \gamma_2=-\left(1+\texttt{sgn}(R)\dfrac{1}{2a^2}\right)=-\left(1+\dfrac{1}{2\sigma_{\alpha}\sigma_n R}\right),\nonumber\\ 
\gamma_{2\infty}&=\lim_{Y\to -\infty (Z\to 0) \text{ in } R>0} \gamma_2=+\infty, \\
\gamma_{2\infty}&=\lim_{Y\to +\infty (Z\to +\infty) \text{ in } R\leqslant0} \gamma_2=-\infty. \nonumber
\end{align} \end{linenomath}

Therefore we can make a conclusion:

\begin{proposition} \label{th:prop2}
A distribution Eq.~\eqref{eq:Z2PDF} of a quantity whose logarithm is described by Eq.~\eqref{eq:Y2} has different asymptotic behavior on both sides of infinity: the asymptotic power-law behavior approximated by the log-CS$\chi_1$ distribution Eq.~\eqref{eq:Z31PDF} on one side and the log-normal behavior Eq.~\eqref{eq:Z41PDF} on the opposite side, whose asymptotic exponents are determined by Eq.~\eqref{eq:gamma2lim}. The choice of sides depends on the sign of the correlation in Eq.~\eqref{eq:R}.
\end{proposition}

This implies an important fact that the geometrically growing system Eq.~\eqref{eq:ansatz} could have the asymptotic power-law behavior coming from the log-CS$\chi_1$ distribution, though which could be alternated by the log-normal one in certain condition. 

The asymptotic power-law is an advantage of the log-CS$\chi_1$. However, its singularity makes it impossible to extend the log-CS$\chi_1$ approximation on the opposite side while the real distribution of the geometrically growing system and ``the correct form'' have the continuity on both sides. This drawback of the log-CS$\chi_1$ can be replenished by the log-normal $Z_4$ extended on that side.

So far we have four approximations to ``the correct form'' of $Y_2$ in Eq.~\eqref{eq:Y2}: ``the analytic approximations''$Y_3$ and $Y_4$ have the same mean and variance with $Y_2$ while ``the asymptotic approximations'' $Y_{31}$ and $Y_{41}$ are consistent with $Y_2$ at infinities. We compare their corresponding $p_Z(Z)$s with the ``correct form'' $p_{Z_2}(Z)$ and ``observed data'' for an instance of the simulation in the previous Sec.~\ref{sec:numsim} where the final spectrum (Fig.~\ref{fig:Configb}) in the mirror-reflection acceleration plays a role of the ``data.'' In order to compare with histogram, the total number of particles and the bin width in histogram should be multiplied to $p_Z(Z)$ for the normalization and integral within the bin width. The parameters $\mu_n, \sigma_n, \mu_{\alpha}, \sigma_{\alpha}, \mu_i, \sigma_i$ and $R$ are extracted from the simulation. $\mu_n, \sigma_n$ are calculated for encounters of $\alpha\geqslant10^{-6}$ (which is the calculation precision). The number of encounters $n$ in the simulation can be replaced with the duration or the age of growth for a continuously growing system.

\begin{figure*}
\centering
\subfigure[]{\includegraphics[width=0.4\textwidth]{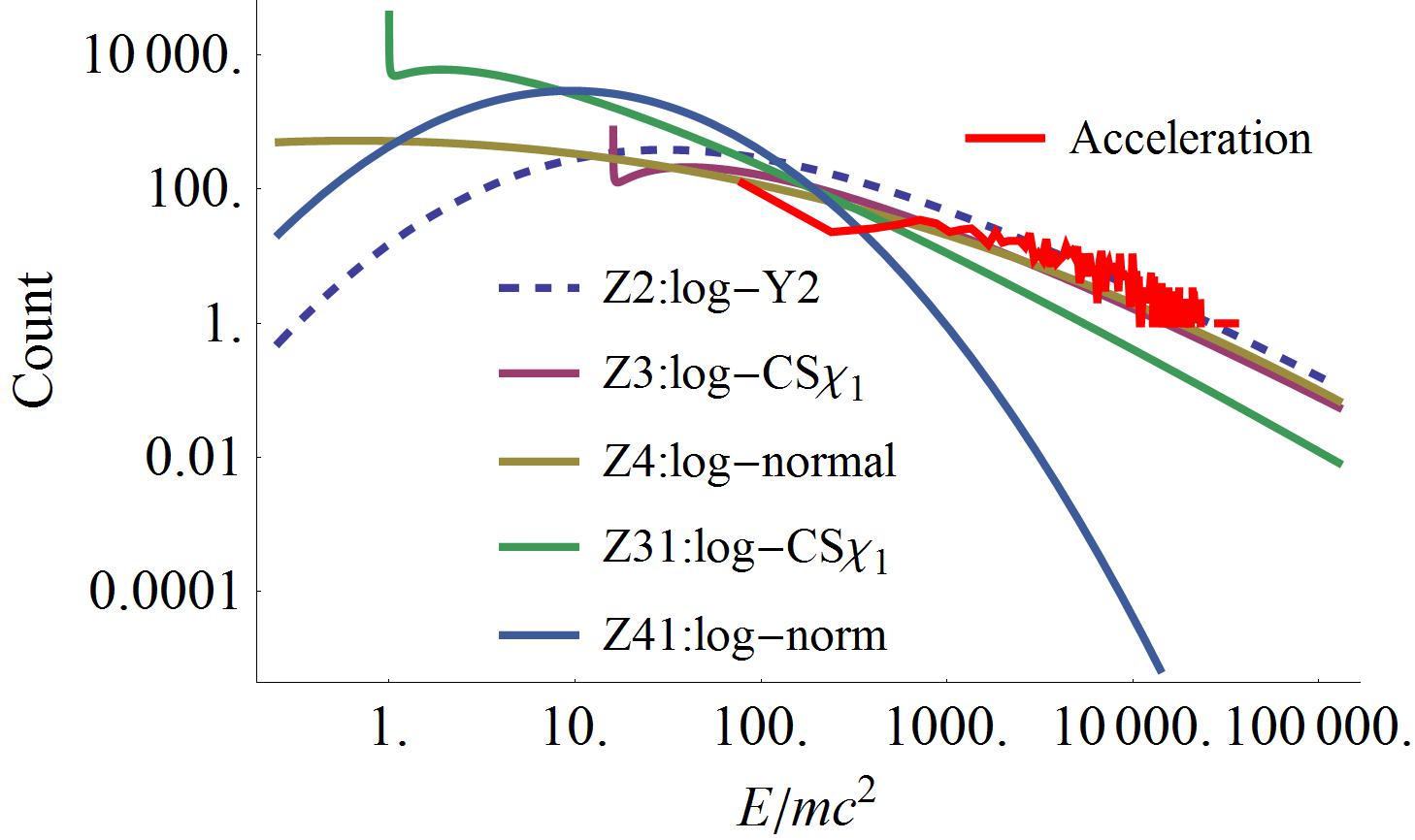}\label{fig:sumcompa}}
\subfigure[]{\includegraphics[width=0.4\textwidth]{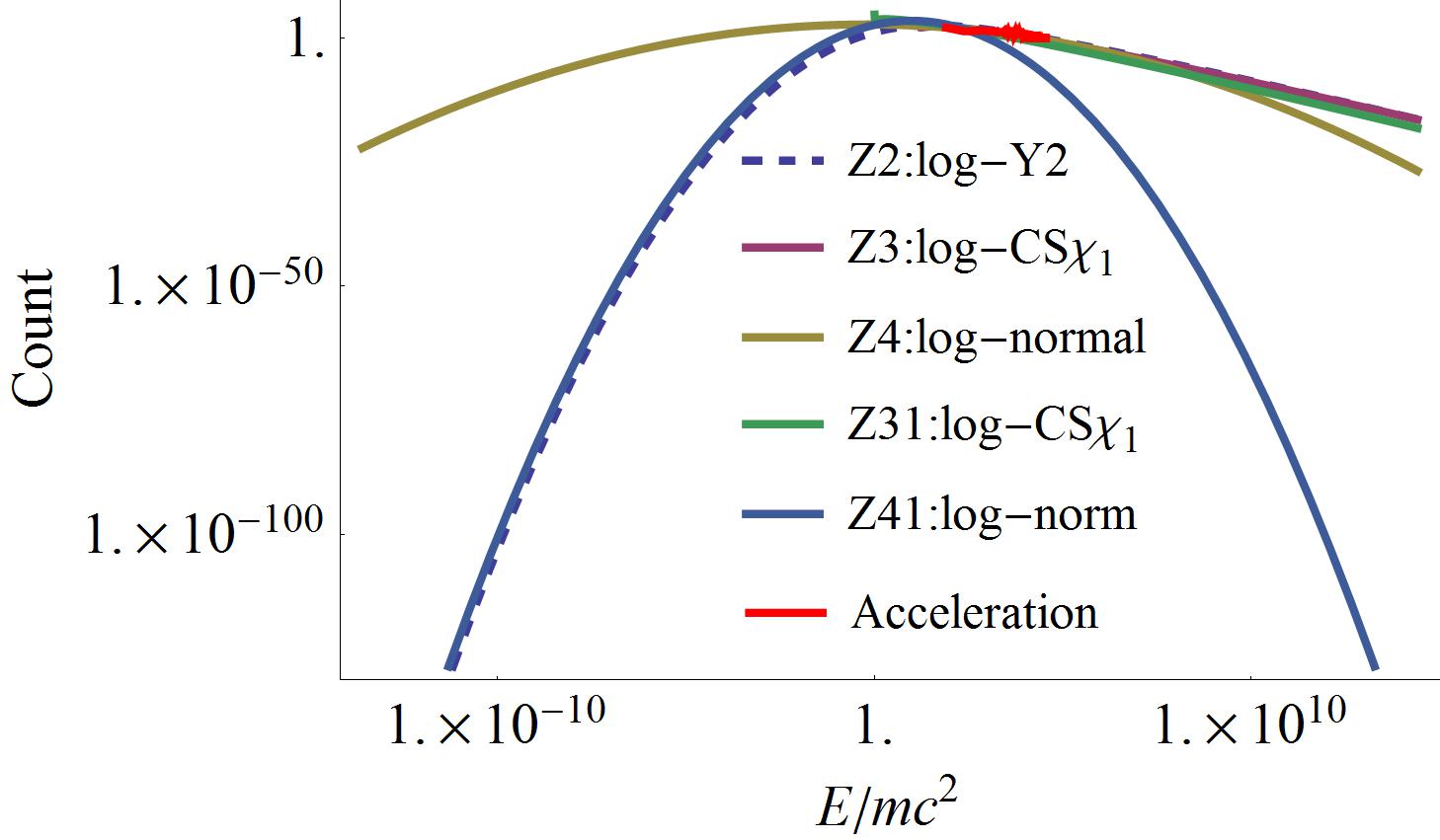}\label{fig:sumcompb}}
\caption{\label{fig:sumcomp} (a) The close-up view for comparison of the final spectrum in the mirror-reflection acceleration (red) with its various approximations $p_Z(Z)s$. Bins are linear-spaced. (b) The remote view for comparison with the various approximations $p_Z(Z)s$.} 
\end{figure*}

Figure~\ref{fig:sumcompa} and \ref{fig:sumcompb} show those comparisons in a close-up and remote views. We can see that at the upper limit the ``observed data'' is consistent with the ``correct form'' even in spite of that the elementary distributions of $n, \alpha$ and $Y_0$ are deviated from normal as aforementioned. This shows that our assumption for the normality of the elementary distributions is plausible. The consistency is attained even at a relatively small number of particles (=1000), which may cause a great randomness. As analytic approximations, $p_{Z_3}(Z)$ and $p_{Z_4}(Z)$ show the closeness to $p_{Z_2}(Z)$ within the scope of ``the observed data'' while $p_{Z_{31}}(Z)$ and $p_{Z_{41}}(Z)$ do not fit to it within that scope. The consistency gets better at the upper limit. However, at the infinities, $p_{Z_{31}}(Z)$, $p_{Z_{41}}(Z)$ and even $p_{Z_{3}}(Z)$ are consistent to $p_{Z_2}(Z)$ while $p_{Z_{4}}(Z)$ shows a great discrepancy. Considering the fact that $Y_3$ and $Y_4$ have the same mean and variance with $Y_2$ and the proposition~\ref{th:prop2}, we can claim:

\begin{proposition} \label{th:prop3}
A distribution Eq.~\eqref{eq:Z2PDF} of a quantity whose logarithm is expressed as Eq.~\eqref{eq:Y2} can be approximate by the log-CS$\chi_1$ distribution Eq.~\eqref{eq:Z3PDF} in the case of positive correlation and by the log-normal Eq.~\eqref{eq:Z4PDF} in the case of negative or zero correlations at least at the upper limit.
\end{proposition}

Do not confuse the upper limit with the infinity mentioned in proposition~\ref{th:prop2}. The upper limit is limited within the finite scope of the data.

We note that $R=0$ corresponds to the log-normal approximation. If the correlation is 0, $Y_2$ can be transformed completely into $Y_4$ whose distribution is normal. In this sense the log-normal can be regarded as a special case of the log-CS$\chi_1$ for $R=0$. So if $R\ll1$, it must be difficult to distinguish the log-normal and log-CS$\chi_1$, i.e., the power-law distribution. This could explain why in certain cases the log-normal and the power-law distributions are indistinguishable.

Comparing $Y_2$ in Eq.~\eqref{eq:Y2} and $Y_3$ in Eq.~\eqref{eq:Y3}, we can find that if $d=0$, that is, the initial distribution of $Y_0=\ln Z_0$ has only one value or is singular, then the distribution of $Y_2$ ($Z_2$) is completely consistent with the CS$\chi_1$ of $Y_3$ (log-CS$\chi_1$ of $Z_3$). This condition looks implausible, however, if we suppose that, for example, every city began with a couple of Adam and Eve, every forest fire broke from a single spark on a tree or every wealth was originated from a penny in the pocket (all of them could imply a fixed initial value, e.g., $Y_0=\ln Z_0=0$), we can regard that many things could embody this condition. That is, any growth of an individual in a group can be so extrapolated to the fixed initial size, which might need the elongation of the age that this condition can be regarded as general. Therefore, we can make another conclusion:

\begin{proposition} \label{th:prop4}
A distribution Eq.~\eqref{eq:Z2PDF} of a quantity whose logarithm is expressed by Eq.~\eqref{eq:Y2} can be approximate uniquely by the log-CS$\chi_1$ distribution Eq.~\eqref{eq:Z3PDF} if the initial distribution is assumed singular.
\end{proposition}

From the proposition~\ref{th:prop1} we can summarize all the above proposition extending to the distribution of the geometrically growing system Eq.~\eqref{eq:ansatz}:

\begin{corollary} \label{th:coro4}
A distribution of the geometrically growing system Eq.~\eqref{eq:ansatz} can be approximated by $p_{Z_{3}}(Z)$ in Eq.~\eqref{eq:Z3PDF} for the positive correlation in Eq.~\eqref{eq:R} and $p_{Z_{4}}(Z)$ in Eq.~\eqref{eq:Z4PDF} for the negative or zero correlation at the upper limit. Besides, the distribution can be approximated by $p_{Z_{3}}(Z)$ in Eq.~\eqref{eq:Z3PDF} if the initial distribution in Eq.~\eqref{eq:normalY0} is singular. The log-CS$\chi_1$ distribution have the asymptotic power-law behavior.
\end{corollary}

Here the upper infinity implies $Z\to+\infty$. We consider only the upper limit but not the upper infinity for the practical cases. The geometrically growing system Eq.~\eqref{eq:ansatz} is supposed to have only the positive size $Z>0$. The log-CS$\chi_1$ approximation can be taken in the cases of the positive correlation or the singular initial distribution while the log-normal approximation can in the cases of the negative or zero correlations.

In practice, on the contrary to the approximation for the normal distribution which covers the infinite domain, $\alpha, n$ and $Y_0$ should be finite, which give a maximum of $Z$ and could make a kind of cutoff at the upper limit. This could give an explanation to commonly appeared truncated power-law spectrum. 

Now, we turn to another problem. We can check the Zipf's law between the log-size and log-rank. \cite{Gabaix1999a} proved that the Zipf's law holds between the log-size of city and its log-rank and the exponent tends toward unity at the upper limit of size. The rank of city is determined by the ordering number of a city in a size-decreasing sequence. So the greatest city has rank 1, the second city has rank 2 and so forth. The rank can be determined by the PDF of size:
\begin{linenomath} \begin{align}\label{eq:rankdef}
r(Z)=N \int_Z^\infty p_{Z}(Z)dZ,
\end{align} \end{linenomath}
where $N$ stands for the number of individuals (e.g. particles or cities). We assume the asymptotic power-law exponent $\gamma_\infty$ at the upper infinity $Z\to +\infty$ (for the log-normal case, $\gamma_\infty$ will be $-\infty$). By making use of Eq.~\eqref{eq:pzpower} we can obtain
\begin{linenomath} \begin{align}
\lim_{Z\to\infty}r(Z)=\frac{C'N}{\gamma_\infty+1} Z^{\gamma_\infty+1},
\end{align} \end{linenomath}
where we should assume $\gamma_\infty+1<0$. So the Zipf's exponent can be written as
\begin{linenomath} \begin{align}\label{eq:zetalim}
\zeta_{\infty}=\frac{d\ln r(Z)}{d \ln Z}=\gamma_\infty+1.
\end{align} \end{linenomath} 

\begin{proposition} \label{th:prop5}
A quantity that has the asymptotic power-law exponent Eq.~\eqref{eq:gamma2lim} for the log-size vs. the logarithm of PDF has also the asymptotic power-law (Zipf's) exponent for its log-size vs. the log-rank as in Eq.~\eqref{eq:zetalim}. 
\end{proposition}

For the case of the asymptotic power-law behavior, the Zipf's exponent can be obtained as
\begin{linenomath} \begin{align}
\zeta_{\infty}=\gamma_\infty+1=-\dfrac{1}{2\sigma_{\alpha}\sigma_n R}.
\end{align} \end{linenomath}
For the asymptotic log-normal behavior, the exponent will be $\pm\infty$.

We have not found here any requirement that the Zipf's exponent should be -1 as indicated in \cite{Gabaix1999a}. We can see that this exponent should be related to the parameters in the specific cases. Perhaps, we have to care that his formula was derived under the assumption of the steadiness of the normalized distribution. There might be another possibility of the Zipf's exponent of -1: the aforementioned $\zeta_\infty$ is a value at infinity but not at the upper limit of finite data. So we have to consider the exponent at the upper limit.

As seen above, we calculate the exponent at the upper limit $z_{max}$ as
\begin{linenomath} \begin{align}
\zeta(z_{max})=\left.\frac{d\ln r(Z)}{d \ln Z}\right|_{Z=z_{max}}=\left.\frac{d r(Z)}{d  Z}\right|_{Z=z_{max}}\frac{z_{max}}{r(z_{max})}=- \frac{N z_{max} p_{Z}(z_{max}) }{r(z_{max})},
\end{align} \end{linenomath}
where Eq.~\eqref{eq:rankdef} is used. We should mention that the upper limit correspond just to the rank 1, i.e. $r(z_{max})=1$. Recall that the multiplication $N p_{Z}(Z) dZ$ gives the frequency (or count) within the bin width $dZ$ around $Z$ in a histogram. The frequency should be also near 1 at the upper limit. If we have $n_b$ log-spaced bins between $Z=z_{min}$ and $Z=z_{max}$ in the histogram such as in Fig~\ref{fig:COVIDHistlog}, the bin width $dZ$ at $z_{max}$ will be
\begin{linenomath} \begin{align}
\left.dZ\right|_{Z=z_{max}}&=z_{max}-z_{min}\left(\frac{z_{max}}{z_{min}}\right)^{(1-\frac{1}{n_b})}=z_{max}-z_{min}\left(\frac{z_{max}}{z_{min}}-\frac{1}{n_b}\left(\frac{z_{max}}{z_{min}}\right)\ln\left(\frac{z_{max}}{z_{min}}\right)-O\left(\frac{1}{n_b^2}\right)\right)\notag\\
&=\frac{z_{max}}{n_b}\left(\ln z_{max}-\ln z_{min}\right).
\end{align} \end{linenomath}
Suppose that
\begin{linenomath} \begin{align}
n_n=\nu\left(\ln z_{max}-\ln z_{min}\right),
\end{align} \end{linenomath}
i.e. the number of bins is $\nu$ times $\ln\left(\frac{z_{max}}{z_{min}}\right)$. We can suppose that the $\nu$ is usually near to unity, i.e. the number of bins is similar to $\ln\left(\frac{z_{max}}{z_{min}}\right)$. Let the frequency within the bin at $z_{max}$ be $h_{max}$: $N p_{Z}(z_{max}) \left.dz\right|_{z=z_{max}}=h_{max}$. Then, we obtain
\begin{linenomath} \begin{align}\label{eq:Zipfexp}
\zeta(z_{max})=\frac{N p_{Z}(z_{max}) z_{max}}{r(z_{max})}=\frac{\nu h_{max}}{r(z_{max})}
\end{align} \end{linenomath}
So we can determine the Zipf's exponent in a certain way:
\begin{proposition}\label{prop6}
The Zipf's exponent between the log-rank and the log-size of a quantity at its upper limit can be determined by the number of log-spaced bins and the frequency at the upper limit bin and the rank at the upper limit as in Eq.~\eqref{eq:Zipfexp}.
\end{proposition}
As mentioned above, considering that $\nu$, $h_{max}$ and even $r(z_{max})$ could be probably unity for a certain distribution including the above approximations, we could expect that the Zipf's exponent at the upper limit should be near unity: $\zeta(z_{max})\approx 1$. This formulation seems applicable to any probability density function which decreases monotonically to zero at the upper infinity. We will see an example in statistics of the COVID-19 pandemic discussed in the next section with this exponent.

\section{Statistics of the COVID-19 pandemic }\label{sec:covid}
We can expect that a pandemic propagation yields the power-law distribution if the epidemiological rule of thumb ``infecting proportional to infected'' preserves. This rule leads to an exponential growth or, in other words, the geometric progress Eq.~\eqref{eq:ansatz}. In fact, however, the lockdown measures or vaccinations and seasonal effects gave very great rises and falls in the history of confirmed cases\footnote{For example, see the website of Our World in Data: \url{https://ourworldindata.org/coronavirus}}, though the infected cases of the COVID-19 are reported to be much more than during the same period last year in the world and the new variants of coronavirus show accelerating propagation. But we can describe the growth by an exponential one if we consider only the first and last numbers of accumulated cases. So we approximate the propagation of the pandemic by the geometric progress. This also could test the applicability of our model to general kinds of growth.

The fact that statistics of the COVID-19 shows the power-law distribution has already been reported. \cite{Blasius2020} computed the distribution of confirmed COVID-19 cases and deaths for countries worldwide to April in 2020 and showed that the distribution follow a truncated power-law over five orders of magnitude. He explained the distribution by both exponential processes in the spread between the countries and the accumulation of the case numbers within each country. This model can be applied only to the early stage of propagation when the number of the propagated countries grows almost exponentially. However, it lasted already more than one year since the outbreak of the pandemic so such a spread between countries was almost saturated. \cite{Beare2020} analyzed the power-law distribution of the pandemic until March in 2020 for counties in US and explained the Pareto slope for the right tail via Gibrat's model for the distributions of the growth rate and, in addition, variance of the propagation age (i.e. the duration of propagation). Gibrat's model supposes that the number of infected is not correlated with the growth rate and our approach also includes this model implicitly. 

We try to explain statistics of the confirmed pandemic cases for 180 countries and areas worldwide\footnote{The data are extracted at WHO coronavirus disease (COVID-19) dashboard: \url{https://covid19.who.int/region/}}, where we ignore the zero-case countries. The summarizing dates are different by countries on between April 21 and May 4 in 2021. These differences are not so serious by generosity of the analysis that this could be included in the variance of the duration of propagation. 

\begin{figure*}
\centering
\begin{tabular}[t]{c}
\subfigure[]{\includegraphics[width=0.2\textwidth]{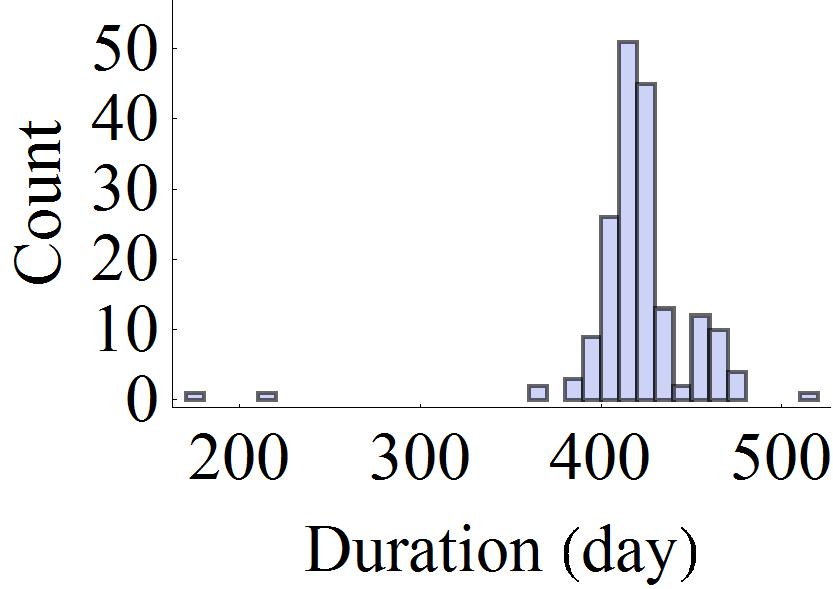}\label{fig:COVIDduratdist}}
\subfigure[]{\includegraphics[width=0.2\textwidth]{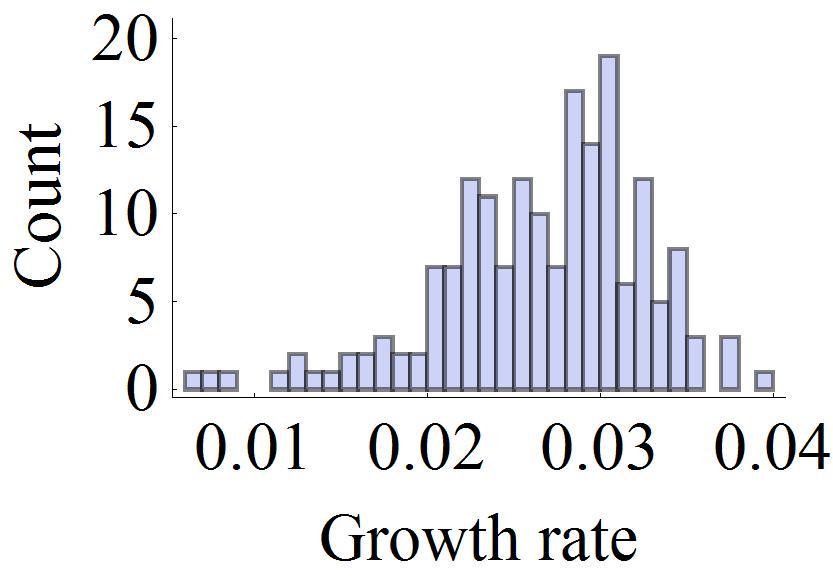}\label{fig:COVIDgratedist}}
\subfigure[]{\includegraphics[width=0.2\textwidth]{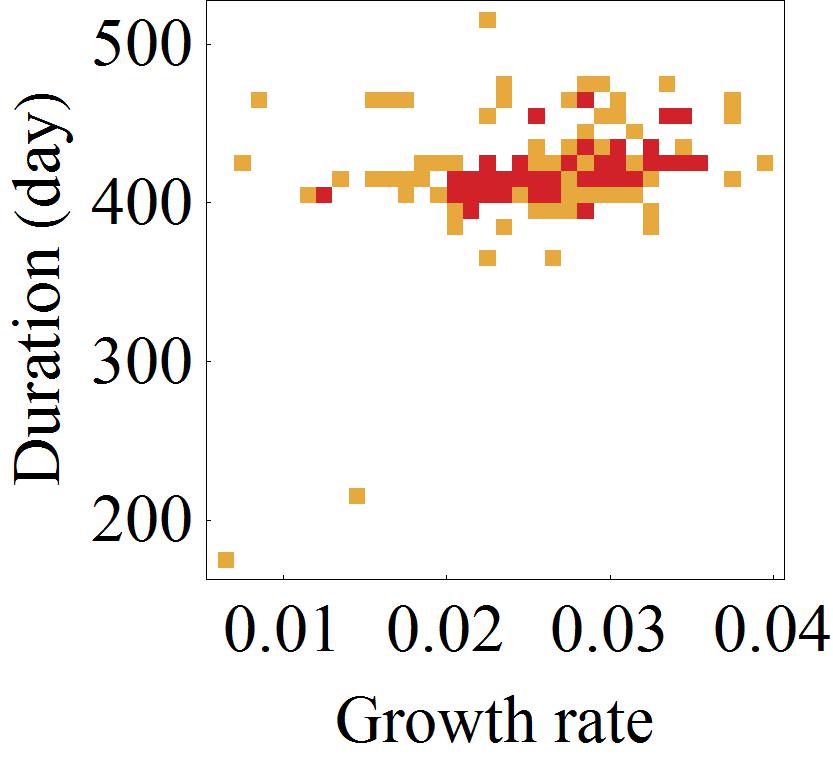}\label{fig:COVIDgrateduratcor}}
\subfigure[]{\includegraphics[width=0.29\textwidth]{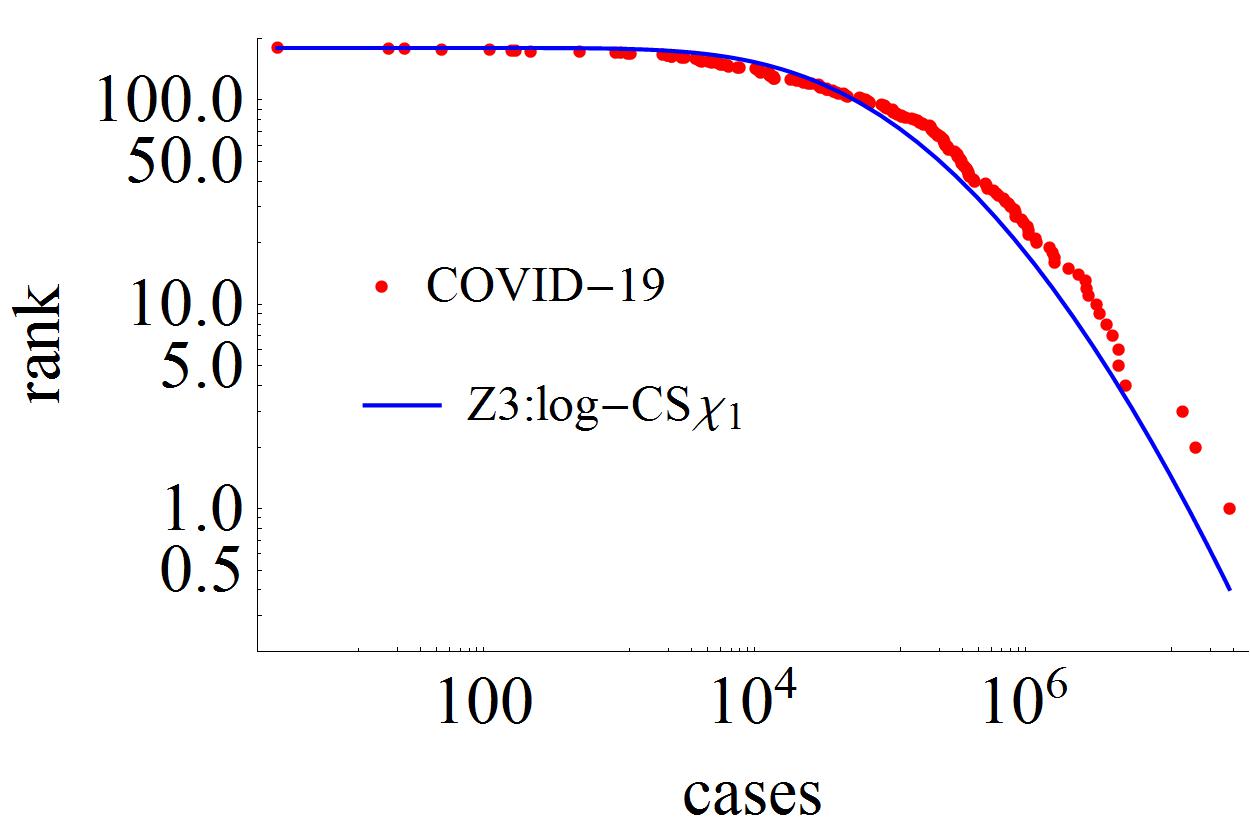}\label{fig:COVIDZipf}}\\
\subfigure[]{\includegraphics[width=0.41\textwidth]{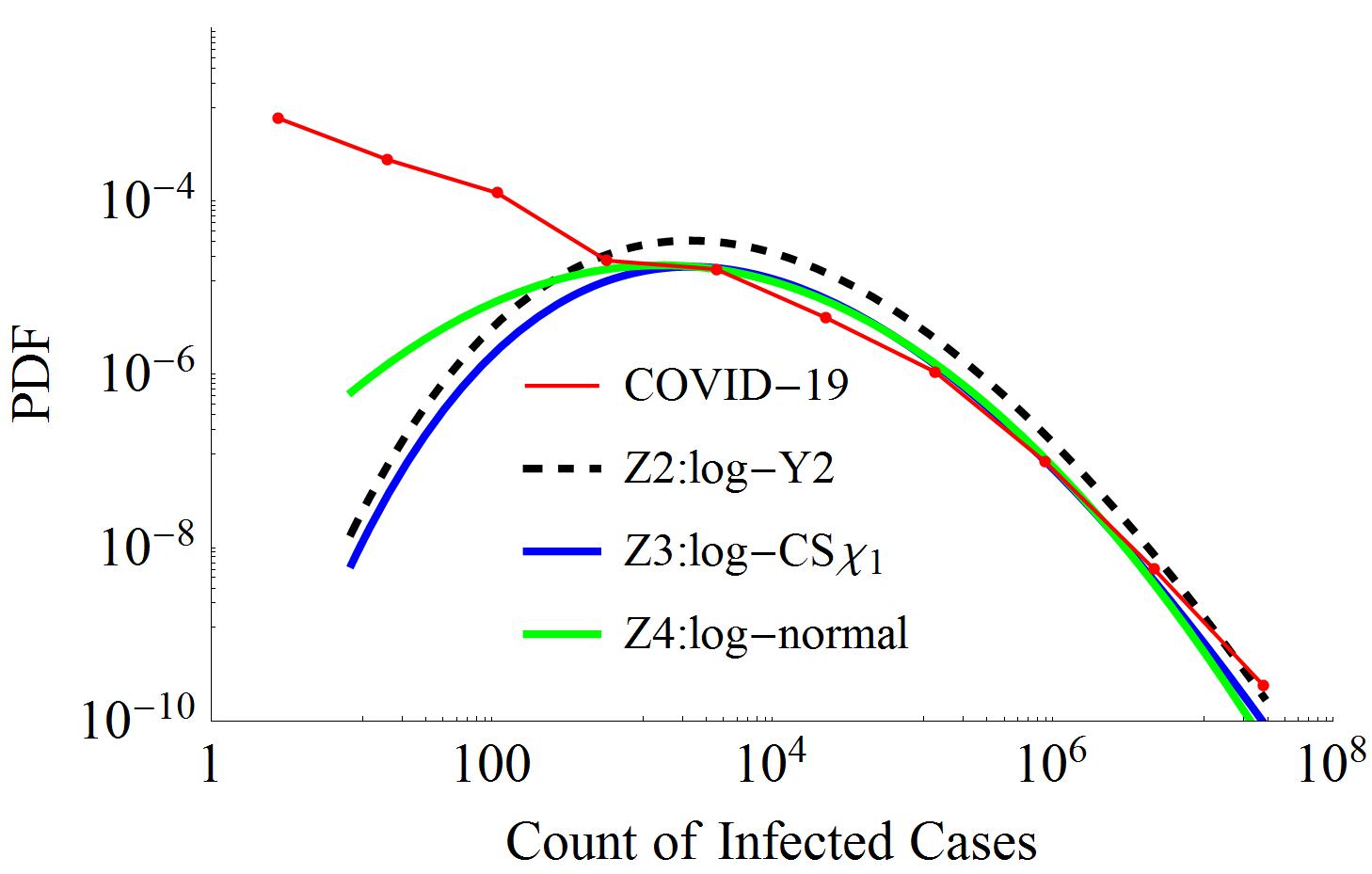}\label{fig:COVIDPDFlog}}
\raisebox{-0ex}[0ex][0ex]{\subfigure[]{\includegraphics[width=0.41\textwidth]{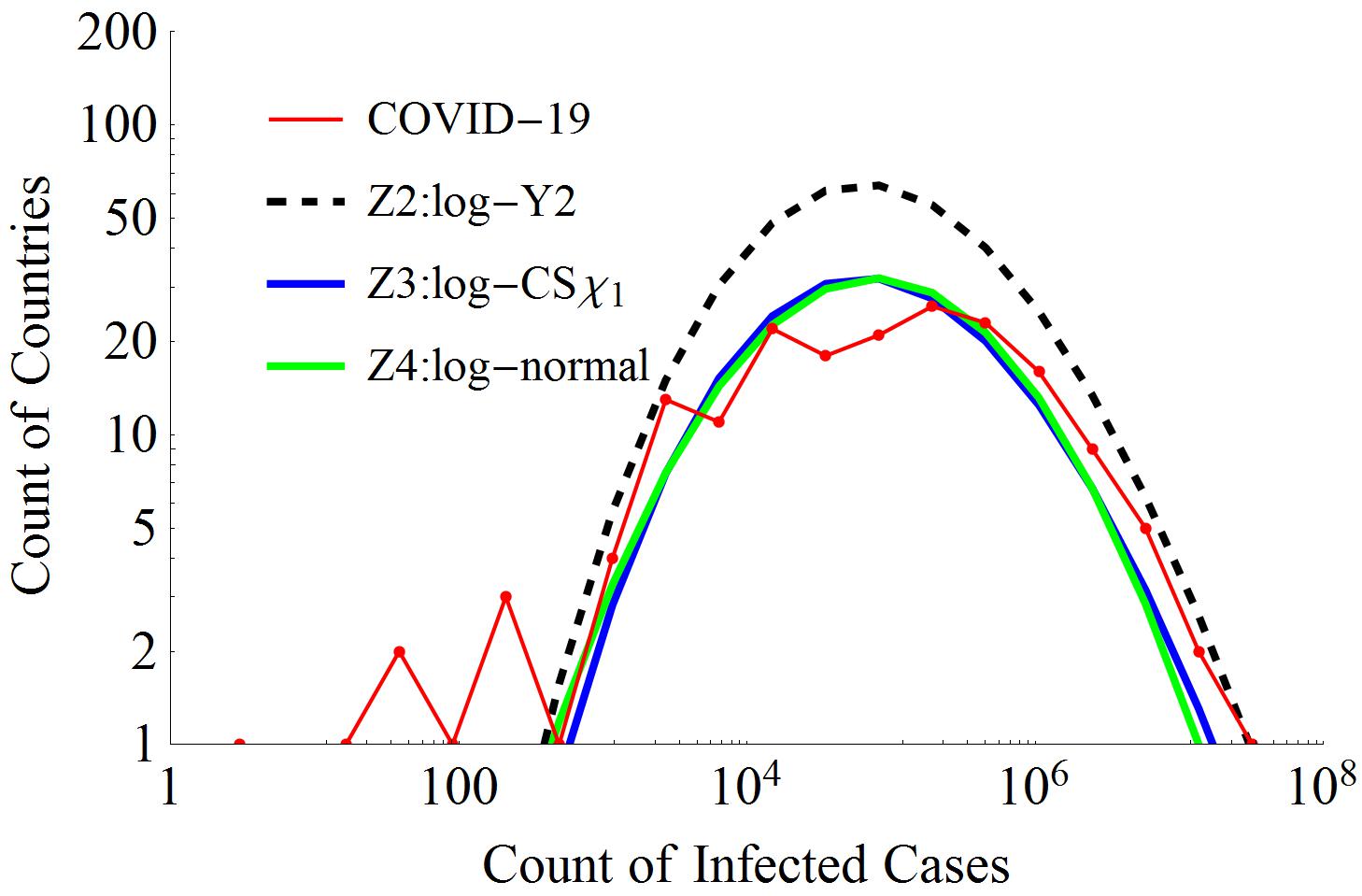}\label{fig:COVIDHistlog}}}
\end{tabular}
\caption{The analysis of the pandemic in May 2021. (a) The histogram of the duration of propagation per country. (b) The histogram of the growth rate. (c) The density pair-histogram between the duration and the growth rate, which shows a correlation between them. (d) The rank of the countries vs. the number of cases in the data (red dot) and log-CS$\chi_1$ approximation (blue curve) on log-log scales. (e) The PDF of the pandemic data (red) and its various approximations. (f) The histogram for the pandemic data (red) and its various approximations. Except lower than 500 cases, the curves seem to be consistent over five orders of magnitude.  In (e) and (f), bins are log-spaced.}
\end{figure*}

We can see that the PDF of the infected cases shows rather a near-power-law form (Fig.~\ref{fig:COVIDPDFlog}) while the histogram is no longer the power-law and appears like a hump (Fig.~\ref{fig:COVIDHistlog}) as a whole. The slope of PDF in log-log diagram varies from -0.6 to -1.6. We can see the convex profile in both diagrams which is typical for our approach.

Here the exponent in Eq.~\eqref{eq:ansatz} is replaced with the duration of propagation in every country. The duration is calculated as difference between the first case date and the summarizing date in days. Even though we take the unit of the duration as a month, the asymptotic slope should not be changed as we can check easily. 

The growth rate is inferred from the number of the first cases, the summarized cumulative cases and the duration of propagation, i.e., the age in each country via Eq.~\eqref{eq:ansatz}. We set the number of the first cases in every country to one. This is an extreme approximation because not only one person brought the pandemic virus into his country. In fact, the number of the reported first cases were 1 in almost countries. However, the analysis can make this deviation ascribed to the growth rate. Therefore we can set $\mu_i=0$ and $\sigma_i=0$. This condition corresponds to the case of singular initial distribution and from the proposition~\ref{th:prop4} lets us consider the log-CS$\chi_1$ approximation to the system. The daily growth rate should be so much smaller than the monthly or yearly ones that the condition of the negligible growth rate is satisfied. 

The Figure~\ref{fig:COVIDduratdist} and ~\ref{fig:COVIDgratedist} show near-normal distributions of the duration and the growth rate in accord with our assumption. Here we obtain $\mu_n$=420, $\sigma_n$=32.5, $\mu_{\alpha}$=0.0266 and $\sigma_{\alpha}$=0.00595. 

A more important parameter is a correlation between the duration and the growth rate. The correlation is small and positive (Fig.~\ref{fig:COVIDgrateduratcor}): $R=0.335$. The positiveness of the correlation could be explained by that a later first case (that is, a shorter duration) could give a warning or preparation to the propagation and lead to a lower growth rate. From the positiveness of the correlation and the proposition~\ref{th:prop3} and ~\ref{th:prop4} in the previous section, we can adopt the log-CS$\chi_1$ distribution as ``the analytic approximation.''

Next, we can estimate the asymptotic slope in Eq.~\eqref{eq:gamma3lim}: $\gamma_{\infty}$=-8.72 for PDF. This is much different from that of observed spectrum of $\sim$-1. Does it mean a fail of the analysis? 

We draw the PDF and histogram for the various approximations with the inferred parameters. Figure~\ref{fig:COVIDPDFlog} and ~\ref{fig:COVIDHistlog} show a perfect coincidence between the observation and the various approximations including ``the correct form'' $p_{Z_2}(Z)$ in Eq.~\eqref{eq:Z2PDF}, the log-CS$\chi_1$ $p_{Z_3}(Z)$ in Eq.~\eqref{eq:Z3PDF} and even the log-normal $p_{Z_4}(Z)$ in Eq.~\eqref{eq:Z4PDF} for more than 500 cases in both the order of magnitude and the slope. The great departure of the asymptotic slope from the slope within the scope of data can be explained in terms of that the asymptotic slope is a value at the infinity which is much far from the upper limit of observed cases. 

The PDFs are deviated from the data for below 500 cases. This could imply that the countries with below 500 cases (they all are equatorial or pacific) are much deviated from the supposed normal distribution of the ensemble of all the countries in either the growth rate or the duration. Really, those countries have the growth rate of $\sim$0.01, which is $\sim$3$\sigma_{\alpha}$ out of the whole ensemble average $\mu_{\alpha}$=0.0266. However, the countries with above 500 cases cover 99.9995\% of all the cases and 95.6\% of all the considered countries and areas. So we have no reason to reject a hypothesis that statistic of the pandemic can be explained by our approach or even follow the log-CS$\chi_1$. We can be sure that the distribution has been evolved from a straight power-law in the early stage of propagation to the current hump shape. In this sense, the our approach can be said to give a dynamic power-law distribution.

We could not find yet a comparable model to reproduce the whole profile of distribution: the Gibrat's models such as \citet{Gabaix1999a} and \cite{Beare2020} give only the exponent of the upper limit while the preferential attachment such as Yule-Simon process or \cite{Barabasi1999} and the model of \cite{Blasius2020}, though can give the whole profile, require an endless propagation to new countries which has been already stopped or saturated for the COVID-19 pandemic in May 2021. Even the self-organized criticality model such as \cite{Bak1987} seems not able to be applied unless the current status of the pandemic can be said to reach the so called critical one. Thus we could compare only the exponent at the upper limit.

We can determine the Zipf's exponent between the log-cases (the logarithm of the number of infected cases) and log-rank (Fig.~\ref{fig:COVIDZipf}). Connecting the uppermost two points, the exponent seems to be $\sim$-1.2 for the data which is at odds with the value of -1 predicted by \cite{Gabaix1999a} though this phenomena could be described by Gibrat's model. However, the value of Zipf's exponent inferred from data is so much liable depending on the selection of points or the approximation method. Thus we infer this exponent for the log-CS$\chi_1$ of $p_{Z_3}(Z)$, which is -1.35. Meanwhile, Eq.~\eqref{eq:Zipfexp} gives -1.46 where $h=0.48, \nu=1.24$ and $r_{z_{max}}=0.40$ which is also calculated for the log-CS$\chi_1$. The asymptotic exponent determined by Eq.~\eqref{eq:zetalim} will be much greater since this value is determined at infinity. Though those values differ a bit, We can see in Fig.~\ref{fig:COVIDZipf} that the Zipf's exponent can be approximated by our approach very well at the upper limit and even all over the scope.

\section{Conclusion and discussion}

In this paper we consider a power-law distribution in the geometrically growing system. The key point here is the assumption for the variance of the number of encounter or the age $n$ and the growth rate $\alpha$ and the correlation between them. They give the $\chi^2$ behavior to reach the asymptotic power-law distribution. In fact, the variance of the growth rate was mentioned in Gibrat's model. Even the variance of the number of encounter or age were mentioned previously. However, if they are independent, we can obtain only the log-normal but not the log-$\chi^2$ or the asymptotic power-law behavior. The above logic shows that the log-normal distribution can be obtained as a special case of the log-CS$\chi^2$ for the correlation of zero.

In the derivation, we made several assumptions: the negligible growth rate, the normal distributions of the growth rate, the age and the logarithm of the initial size and the correlation between the growth rate and the age. As we have seen in Sec.~\ref{sec:covid}, though the monthly growth rate might be significant, the daily or even hourly growth rates could be negligible. So we think that the assumption of negligible growth rate naturally holds. And we have seen that the correlation between the growth rate and the age can be given even in arbitrary cases when there is no systematic cause. So the first and last assumptions can be seen non-artificial or natural. The normality of distribution seems some artificial though statistically plausible. Though the above examples showed good approximations in spite of that the distributions of parameters deviate from the normal ones, we have to investigate the effects of those deviations on the resultant distribution. Obviously, the finiteness and discreteness of the parameters will cause a cutoff and great randomness in the upper or lower limits. The various distributions of parameters should be inspected.

A greatest advantage of the mechanism here might be a comprehensiveness. The slope of spectrum seems to depend explicitly on the global statistics but implicitly on the local physics or microeconomics and even hardly on the initial distribution. This could explain the power-law distribution only through their statistics but without considering local physics or microeconomics. We can not make any assumption on the local circumstances which might be turned out impractical. Less dependency on the local physics or microeconomics is one of important advantages to support this theory.

Furthermore, any slope of the power-law spectrum can be explained. And the mechanism seems to be able to reproduce all kinds of the indistinguishable power-law and log-normal depending only on the sign of correlation and even power-law with a cutoff. As we have seen above, the analysis have a generosity: the different summarizing time for each individual can be embraced into the age. So the power-law distribution can be given at any profile of time cross-section of the evolution. And the mechanism can be applicable even in the case where the preferential attachment can not be applied, e.g., the saturated propagation between countries of the pandemic.

Another advantage may be the ability to analyze the dynamically evolving system. As we can see, not only the straight power-law (though which is not reproduced in our analysis) in the early stage but also the hump-like distribution in the old stage can be explained in the same theory. The creation of the new individual such as new genus, new infected country, new city and so on, can be considered in the distribution of the age on the statistical way without introducing any artificial or specific process.

We can find that this mechanism includes many peculiar properties in the previous works though in different formats: the compatibility of the power-law with the log-normal, an asymptotic power-law behavior, the variances of time and growth rate, a correlation such as ``the first-mover advantage'' and so on. The mechanism can explain ``the roll-over,'' that is a decrease of lower part appeared in most practical power-law distributions, only on a statistical way.

The analysis can be used to identify the elementary processes such as acceleration and growth. Estimating parameters, we can infer the local process. However, a problem is still remained: 7 parameters $\mu_n, \sigma_n, \mu_{\alpha}, \sigma_{\alpha}, \mu_i, \sigma_i\text{ and } R$ are degenerated to only 3 parameters $A, B$ and $C$ or 2 parameters $F$ and $G$. This can be resolved by additional analysis on local physics or microeconomics. 

The formalism here is general, simple and obvious. Known the history of progress, the elementary parameters can be evaluated directly from that without a best-fitting from a resultant distribution. There is no fine-tuned or unquantifiable parameter. The slope is so dependent on the variances of the parameters that could be labile by ensemble or evolution. Many things including the population in city, the wealth distribution and the number of scientific citations could be explained by the analysis. The continuously progressing world will show much more phenomena of the power-law distribution. The power-law distribution or, exactly speaking, the log-CS$\chi_1$ including the log-normal distribution as its specific case seem to be one of inborn properties of the nature.

\section*{Acknowledgements}

K. Chol-jun thanks prof. Kim Jik-su for continuous discussion and supporting.

\section*{Conflict of interest}
The author has no conflicts to disclose.

\section*{Data availability}
Data used in this paper are available at the website addresses indicated or by corresponding with the author.

\section*{Notes}
\setcounter{equation}{0}
\setcounter{figure}{0}

\begin{note}\label{note:CRaccelintro}
The Fermi acceleration is a powerful candidate mechanism for acceleration of the cosmic ray particles to explain the power-law spectrum of their energy. The first-order Fermi acceleration at the shock front and the second-order Fermi acceleration through the diffusive cloud are well known. You can see details of the Fermi acceleration, for example, in \citet{Gaisser1990}. At each encounter the average increment of the particle energy $E$ is expressed by
\begin{linenomath} \begin{align}
M\left[\frac{\Delta E}{E}\right] \approx \frac{4}{3}\beta \qquad &\text{for the first-order Fermi acceleration}, \notag\\
M\left[\frac{\Delta E}{E}\right] \approx \frac{4}{3}\beta^2 \qquad &\text{for the second-order Fermi acceleration,} \notag
\end{align} \end{linenomath}
where $\beta=v/c$ and $v$ stands for the velocity of the shock front in the fist-order mode and for the velocity of the cloud in the second-order mode. The $M[\cdot]$ implies the mean value.

Note that the Fermi acceleration gives also the geometric or exponential growth of the particle energy just as the preferential attachment though their mechanisms are different.
\end{note}

\begin{note}\label{note:lognormal}
A log-normal distribution is the distribution of a quantity whose logarithm follows the normal distribution.  
\end{note}

\begin{note}\label{note:normdistsup}
The assumption for the normal distribution of $\ln Z_0$ means in turn that $Z_0$ is log-normally distributed. If we suppose that an initial state of the system might have passed some pre-history before the ``essential'' geometric growing stage, the initial distribution should be already close to the log-normal by the multiplicative central limit theorem. Anyhow, we can approximate the distribution of $\ln Z_0$ by the normal distribution without loss of generality. 
\end{note}

\begin{note}\label{note:alphancor}
The correlation between the growth rate $\alpha$ and the growth number $n$ introduced here can be dubbed a ``secondary preferential attachment'' if we can generalize the preferential attachment to imply a systematic trends affecting on growth in the geometrically growing system itself. In fact, \cite{Yule1924} and \cite{Simon1955} have used the preferential attachment in the allocation of a newly born individual to lead the geometrically growing system. The correlation seems to reproduce also ``the first-mover advantage'' \citep{Newman2009} though the formalism differs. They express these words in terms of a probability in step-by-step process but not a correlation like here. An anti-correlation, i.e., a negative correlation is possible. For example, there is a folk idiom ``the last-mover advantage'' in our country. The correlation can be given even arbitrarily without any systematic cause, which of course could not be estimated a priori.
\end{note}

\begin{note}\label{note:orthsum}
We took into account of that a sum $x_3$ of two independent normal-distributed variables $x_1$ in $N(\mu_1,\sigma_1)$ and $x_2$ in $N(\mu_2,\sigma_2)$ follows also a normal distribution $N(\mu_1+\mu_2,\sqrt{\sigma_1^2+\sigma_2^2})$. 
\end{note}

\end{document}